\documentclass[12pt,a4paper]{article}

\usepackage{amsmath}
\usepackage{amssymb}
\usepackage[nosort]{cite}

\makeatletter
\@addtoreset{equation}{section}
\makeatother

\makeatletter
\let\old@startsection=\@startsection
\renewcommand{\@startsection}[6]{\old@startsection{#1}{#2}{#3}{#4}{#5}{#6\mathversion{bold}}}
\makeatother

\newcommand{\hypref}[2]{\ifx\href\asklfhas #2\else\href{#1}{#2}\fi}

\let\oldPhi=\Phi
\let\oldPsi=\Psi
\let\oldGamma=\Gamma
\let\oldDelta=\Delta
\let\oldSigma=\Sigma
\let\oldTheta=\Theta
\renewcommand{\Phi}{\mathnormal{\oldPhi}}
\renewcommand{\Psi}{\mathnormal{\oldPsi}}
\renewcommand{\Gamma}{\mathnormal{\oldGamma}}
\renewcommand{\Sigma}{\mathnormal{\oldSigma}}
\renewcommand{\Delta}{\mathrm{\oldDelta}}
\renewcommand{\Theta}{\mathnormal{\oldTheta}}

\newcommand{\sfrac}[2]{{\textstyle\frac{#1}{#2}}}
\newcommand{\half}{\sfrac{1}{2}}
\newcommand{\quarter}{\sfrac{1}{4}}

\newcommand{\grSU}{\mathrm{SU}}
\newcommand{\grSL}{\mathrm{SL}}

\newcommand{\grSO}{\mathrm{SO}}
\newcommand{\order}[1]{\mathcal{O}(#1)}
\newcommand{\eps}{\varepsilon}
\newcommand{\Lagr}{\mathcal{L}}
\newcommand{\superN}{\mathcal{N}}
\newcommand{\gym}{g_{\scriptscriptstyle\mathrm{YM}}}
\newcommand{\gtwo}{g_2}
\newcommand{\diag}{\mathop{\mathrm{diag}}}
\newcommand{\Tr}{\mathop{\mathrm{Tr}}}

\newcommand{\cdott}{\mathord{\cdot}}

\newcommand{\trans}{{\scriptscriptstyle\mathsf{T}}}
\newcommand{\conj}{\ast}

\newcommand{\irrep}[1]{{\mathbf{#1}}}

\newcommand{\Real}{\mathbb{R}}
\newcommand{\Comp}{\mathbb{C}}
\newcommand{\cder}{\mathcal{D}}

\newcommand{\OpB}{\mathcal{O}}
\newcommand{\OpV}{\mathcal{Q}}

\newcommand{\lrbrk}[1]{\left(#1\right)}
\newcommand{\bigbrk}[1]{\bigl(#1\bigr)}

\newcommand{\bigvev}[1]{\bigl\langle#1\bigr\rangle}

\newcommand{\comm}[2]{[#1,#2]}
\newcommand{\acomm}[2]{\{#1,#2\}}

\newcommand{\bigeval}[1]{#1\big|}
\newcommand{\lreval}[1]{\left.#1\right|}

\newcommand{\nln}{\nonumber\\}
\newcommand{\nl}{\nonumber\\&&\mathord{}}
\newcommand{\nlnum}{\\\nonumber&&\mathord{}}

\newcommand{\eq}{\mathrel{}&=&\mathrel{}}
\newenvironment{myeqnarray}{\arraycolsep0pt\begin{eqnarray}}{\end{eqnarray}\ignorespacesafterend}
\newenvironment{myeqnarray*}{\arraycolsep0pt\begin{eqnarray*}}{\end{eqnarray*}\ignorespacesafterend}

\let\displaymath=\[
\let\enddisplaymath=\]
\def\[{\begin{equation}}
\def\]{\end{equation}}
\def\<{\begin{myeqnarray}}
\def\>{\end{myeqnarray}}

\begin{document}

\thispagestyle{empty}
\begin{flushright}\footnotesize
\texttt{hep-th/0211032}\\
\texttt{AEI 2002-089}
\end{flushright}
\vspace{2cm}

\begin{center}
{\Large\textbf{BMN Operators\\and\\Superconformal Symmetry}\par}
\vspace{2cm}

\textsc{Niklas Beisert}
\vspace{5mm}

\textit{Max-Planck-Institut f\"ur Gravitationsphysik\\
Albert-Einstein-Institut\\
Am M\"uhlenberg 1, D-14476 Golm, Germany}
\vspace{3mm}

\texttt{nbeisert@aei.mpg.de}\par\vspace{3cm}

\textbf{Abstract}\vspace{7mm}

\begin{minipage}{14cm}
Implications of $\superN=4$ superconformal symmetry 
on Berenstein-Maldacena-Nastase (BMN) operators with two
charge defects are studied both at finite charge $J$ and in the BMN limit. 
We find that all of these belong to a single 
long supermultiplet explaining a 
recently discovered degeneracy of
anomalous dimensions on the sphere and torus.
The lowest dimensional component is 
an operator of naive dimension $J+2$ transforming in 
the $[0,J,0]$ representation of $\grSU(4)$.
We thus find that the BMN operators are
large $J$ generalisations of the Konishi operator at $J=0$.
We explicitly construct descendant operators by supersymmetry 
transformations and
investigate their three-point functions 
using superconformal symmetry.
\end{minipage}

\end{center}

\newpage
\setcounter{page}{1}
\setcounter{footnote}{0}

%%%%%%%%%%%%%%%%%%%%%%%%%%%%%%%%%%%%%%%%%%%%%%%%%%%%%%%%%%%%%%%%%%%%%%%%%%%%%%%%%%%%
%%%%%%%%%%%%%%%%%%%%%%%%%%%%%%%%%%%%%%%%%%%%%%%%%%%%%%%%%%%%%%%%%%%%%%%%%%%%%
\section{Introduction and overview}
\label{sec:intro}

In their insightful investigation of strings on a plane-wave background 
\cite{Berenstein:2002jq}
Berenstein, Maldacena and Nastase (BMN)
proposed a method to construct 
$\superN=4$ $\grSU(N)$ super Yang-Mills theory
operators dual to these string states. 
These so-called BMN operators are single-trace operators with
near extremal charge $J$ under an $\grSO(2)$ subgroup of the
$\grSO(6)\sim \grSU(4)$ R-symmetry. They are obtained by 
inserting a few impurities into the trace of an operator with
extremal $\grSO(2)$ charge. On the string theory side
the extremal operator corresponds to
a string in the ground state and the impurities 
correspond to (localised) string excitations. 
After a Fourier-mode decomposition 
of the positions of defects
the field theory operators 
were seen to have finite anomalous dimension 
in the proposed limit
\[\label{eq:BMNlimit}
N,J\to \infty\quad\mbox{with}\quad
\lambda'=\frac{\gym^2 N}{J^2}\quad\mbox{and}\quad
\gtwo=\frac{J^2}{N}\quad\mbox{fixed},
\]
the BMN limit. On the string theory side 
this corresponds to the fact that the plane-wave background 
is a Penrose limit of $\mathrm{AdS}_5\times \mathrm{S}^5$
\cite{Blau:2002dy}.
It was suggested and shown to leading order that 
the spectrum of scaling dimensions of these operators matches
the light-cone energy spectrum of string theory.

This inspiring proposal has sparked numerous works on
both the string theory and field theory side. 
On the string theory side the interest 
\cite{Metsaev:2001bj,Spradlin:2002ar,Gopakumar:2002dq,Lee:2002rm,
Spradlin:2002rv,Schwarz:2002bc,Pankiewicz:2002gs,Pankiewicz:2002tg}
is due to major simplifications of the calculation 
in the plane-wave limit.
For the first time there is an opportunity to compare
directly string states with field theory and
test the AdS/CFT conjecture beyond supergravity. 
On the gauge theory side it has given rise to a number
of investigations 
\cite{Kristjansen:2002bb,Gross:2002su,Constable:2002hw,Arutyunov:2002xd,Bianchi:2002rw,
Chu:2002pd,Santambrogio:2002sb,Huang:2002yt,Parnachev:2002kk,Gursoy:2002yy,
Beisert:2002bb,Constable:2002vq,Eynard:2002df,Janik:2002bd,Klose:2003tw}
which also brought about new insights on $\superN=4$ SYM. 
Some works try to match
string theory to field theory
\cite{Chu:2002pd,Parnachev:2002kk,Verlinde:2002ig,
Gross:2002mh,Vaman:2002ka,Pearson:2002zs,Gomis:2002wi}.
\medskip

In \cite{Kristjansen:2002bb,Constable:2002hw} it was noticed
that the BMN limit is very different from the usual 
`t Hooft limit in that it permits all-genus
amplitudes and not just planar ones. 
It involves two meaningful parameters, $\lambda'$,
the coupling constant, and 
$\gtwo$, an effective genus counting parameter.
The corrections to two-point correlation functions
on the torus, $\order{\gtwo^2}$, and at one-loop,
$\order{\lambda'}$, were calculated in these articles.
To compute the torus correction to the anomalous dimension
all operators with common quantum numbers
have to be redefined (mixed) in order to 
normalise and diagonalise their two-point 
functions. 
A crucial insight was that not only the
original, single-trace BMN operators, 
but also similar multi-trace operators have
to be taken into account \cite{Bianchi:2002rw}.
Including these in the diagonalisation procedure 
the anomalous dimension on the torus was found
to be \cite{Beisert:2002bb,Constable:2002vq}
\[\label{eq:BMNdimtorus}
\Delta^J_{n}\sim J+2+\frac{\lambda'}{8\pi^2}
\lrbrk{8\pi^2 n^2+g_2^2\lrbrk{\frac{1}{6}+\frac{35}{16\pi^2 n^2}}}.
\]
In \cite{Bianchi:2002rw,Beisert:2002bb,Constable:2002vq}
it was furthermore noticed that there are
different flavours of BMN operators with two 
scalar impurities. 
With respect to the residual $\grSO(4)$ subgroup of $\grSO(6)$,
which is orthogonal to the $\grSO(2)$ charge subgroup,
a scalar impurity transforms in the $\irrep{4}$ irreducible representation.
Two scalar impurities therefore correspond to the tensor product 
$\irrep{4}\times \irrep{4}=\irrep{1}+\irrep{3}+\irrep{\bar 3}+\irrep{9}$.
On general grounds one should assume that operators transforming 
in different irreducible representations have different properties.
Nevertheless, it was observed that all $16$ operators 
have the \emph{same} anomalous dimension even on the torus.
This is a clear hint at an enlarged symmetry that relates the operators.
In a response to the preprints of \cite{Beisert:2002bb,Constable:2002vq}, 
J. Maldacena and M. van Raamsdonk
kindly pointed out that this degeneracy should be due to a 
symmetry that lies at the heart of $\superN=4$ SYM, 
\emph{super}symmetry%
\footnote{
This was also noted in \cite{Berenstein:2002jq} and a 
revised version of \cite{Constable:2002vq}.}. 
One purpose of this work is to make this 
statement more explicit and work out exactly
how supersymmetry relates these and other operators. 
A first step in this direction was 
already taken in \cite{Gursoy:2002yy}
where the vector-scalar BMN operator was related to
the two-scalar BMN operators by means of supersymmetry.
\medskip

Supersymmetry is not the full symmetry group of 
$\superN=4$ SYM. As a massless field
theory it also exhibits conformal symmetry. 
Due to non-renormalisation of the coupling constant 
in this special theory conformal symmetry is
not broken by quantum corrections. Moreover, in combination with supersymmetry
it enhances to \emph{superconformal} symmetry. 
Like conformal symmetry, superconformal symmetry 
puts severe constraints on the correlation functions
of the theory. 
A special role is played by the two-point and
three-point functions.
Two-point functions are uniquely fixed
and contain information about the
conformal dimension of an operator.
On the string theory side the anomalous dimension 
corresponds to the light-cone energy.
In this context superconformal symmetry has proven useful for
the exact determination of the planar anomalous dimension of 
BMN-like operators \cite{Santambrogio:2002sb}.
Three-point functions, which are not fully determined,
contain the structure constants for the 
operator product expansion (OPE).
A corresponding structure for these on the string theory side,
if it exists at all, is not yet known.
Nevertheless, as three-point functions are well-defined observables
of field theory, some of them have been explicitly computed
\cite{Chu:2002pd,Beisert:2002bb,Constable:2002vq}.
Out of the six three-point functions stated in 
\cite{Beisert:2002bb}, two vanish 
and three are very similar to one another. 
We will explain the reason for this in terms
of superconformal symmetry.
\medskip

Beyond that little is known about the nature of the 
BMN operators. BMN have provided a heuristic method to construct 
them, but it is unclear how they are distinguished.
It is also not known whether the operators are well-defined in
terms of a $\superN=4$ superconformal theory \cite{Arutyunov:2002xd}
or if the BMN limit is well-defined at all
\cite{Arutyunov:2002xd,Beisert:2002bb}.
We will answer the first question in
terms of representation theory, namely that
the BMN operators with two charge defects 
form multiplets of
$\superN=4$ superconformal symmetry.
The primary operator of the multiplet is the $\grSO(4)$ singlet 
two-scalar operator discussed in \cite{Beisert:2002bb}. 
It transforms under the internal $\grSO(6)$ symmetry 
group as $[0,J,0]$ and has naive scaling dimension $J+2$.
All other BMN operators with two charge defects are descendants.
For instance, the antisymmetric and symmetric-traceless 
two-scalar operators of \cite{Beisert:2002bb} are
level $2$ and $4$ descendants, respectively.
This alternative definition of BMN operators may 
eventually lead to a better understanding
of the BMN limit and an answer to the question whether it 
is a good limit of $\superN=4$ SYM.
\medskip

For the primary operators
we find the explicit form 
\[\label{eq:PrimaryOp}
\OpB^J_n=
\frac{N_0^{-J-2}}{2\sqrt{J+3}}
\sum_{p=0}^J\sum_{m=1}^6\cos\frac{\pi n(2p+3)}{J+3}
\Tr \Phi_m Z^p \Phi_m Z^{J-p},\quad
0<n<\frac{J+3}{2},
\]
at one-loop and at the planar level.
The phase factors are determined such that the operators are fully 
orthonormal at this order even at \emph{finite charge} $J$.
From this expression we derive the form of all bosonic operators by supersymmetry. 

Having resolved the form of the exact phase factors at finite $J$
it is natural to investigate these generalised operators
at finite $J$. They form a sector of operators being nearly protected.
Its large charge limit describes strings
in the plane-wave background. In the low
charge regime we will discover the Konishi operator
($J=0$, $n=1$)
and a couple of dimension four operators 
which have been investigated in the recent years.
We also find an expression for 
the one-loop planar scaling dimension of the operators
which interpolates smoothly between the two regimes
\<
\Delta^J_n=J+2+\frac{\gym^2 N}{8\pi^2}\,8\sin^2\frac{\pi n}{J+3}.
\>
For $J=0$, $n=1$ it reproduces precisely the Konishi anomalous 
dimension $3\gym^2 N/4\pi^2$ and 
in the BMN limit it approaches $J+2+\lambda' n^2$. 
\bigskip

This paper is organised as follows. 
In Sec.~\ref{sec:BMNmulti} we will 
use counting arguments to demonstrate
how BMN operators fill out multiplets of 
$\superN=4$ SYM. This will lead to
an alternative and more general definition of BMN operators.
In Sec.~\ref{sec:form} we determine the form of 
these operators in terms of fields.
It will be seen that the operators are meaningful objects
even at finite charge. We find an expression for their
anomalous dimensions which, in particular, interpolates between
the BMN limit and the Konishi operator.  
In Sec.~\ref{sec:susy} we will work out the expressions
for the descendant operators by 
supersymmetry transformations.
This makes the observed degeneracy of anomalous dimensions manifest
and is shown to yield non-trivial relations among their mixing matrices.
Superconformally covariant correlation 
functions involving the BMN operators
are considered in Sec.~\ref{sec:corr}.
We will show how this leads to relations between 
three-point functions.
In Sec.~\ref{sec:Concl} we 
discuss the results of the preceding sections 
and draw conclusions. 
This leads us to a classification of 
operators in $\superN=4$ SYM 
inspired by, but not limited to the BMN limit.

%%%%%%%%%%%%%%%%%%%%%%%%%%%%%%%%%%%%%%%%%%%%%%%%%%%%%%%%%%%%%%%%%%%%%%%%%
\section{BMN Multiplets}\label{sec:BMNmulti}

Berenstein, Maldacena and Nastase have introduced the 
operators \cite{Berenstein:2002jq}
\[
\sum_{p=1}^J e^{2\pi i np/J} \Tr \phi_i Z^p \phi_j Z^{J-p}
\]
of $\superN=4$ SYM and suggested that they 
behave nicely in the limit 
\eqref{eq:BMNlimit}.
They have since been called `BMN operators'
\footnote{
Operators with more than two insertions
in the string of $Z$s have also been proposed. 
We do not consider these throughout the work
and will always refer to operators
with two impurities.}.
They belong to the class of operators with two defects. 
In this section we will work out the multiplet structure 
of such single-trace operators with no more than two 
charge defects. This will clarify the nature of 
these operators in terms of representation theory of $\superN=4$ SYM. 

A charge defect is explained as follows.
The internal symmetry group $\grSO(6)$ is 
split into $\grSO(2)\times \grSO(4)$.
The defect charge is the difference between 
the naive scaling dimension of an operator and its $\grSO(2)$ charge.
There are six scalars in 
$\superN=4$ SYM transforming in the vector 
representation of $\grSO(6)$.
They will be denoted by
$Z=\sfrac{1}{\sqrt{2}}(\Phi_5+i\Phi_6)$, $\phi_{1,2,3,4}=\Phi_{1,2,3,4}$ 
and $\bar Z=\sfrac{1}{\sqrt{2}}(\Phi_5-i\Phi_6)$.
They carry charge $+$, $0$, $-$ and transform under
$\grSO(4)$ as $\irrep{1}$, $\irrep{4}$, $\irrep{1}$, respectively.
Therefore, the only scalar without a charge defect is $Z$. 

To enumerate the operators,
it is convenient to use 
Young tableau notation of $\grSO(6)$ representations
instead of the more common Dynkin labels notation 
$[a,b,c]$.
An $\grSO(6)$ Young tableau consists of three 
horizontal lines of boxes with decreasing 
size. We will denote such a Young tableau by $(a,b,c)$
with $a\geq b\geq |c|$ 
\footnote{
Representations with $c>0$ are chiral, they
have anti-chiral partners with $c<0$.}.
It corresponds to the Dynkin labels $[b+c,a-b,b-c]$.
Young tableaux are useful in this context because
the maximum $\grSO(2)$ charge in a representation
can be read off directly
as the number of boxes in the first row, $a$. 
Furthermore, the weights with maximum charge 
form a representation under the transverse $\grSO(4)$ 
given by the lower two rows, $(b,c)$.

First we consider single-trace operators made up of $\Delta$ scalars 
with charge $J\geq \Delta-2$, i.e. no
more than two charge defects.
The scalars transform under the fundamental representation
$(1,0,0)$ of $\grSO(6)$, a single box.
The product of $\Delta$ scalars is thus a
sum of representations with Young tableaux of 
$\Delta$, $\Delta-2$, $\Delta-4$, \ldots boxes. 
As we are interested only in operators with at most two
charge defects, $J\geq \Delta-2$, and the charge in
a representation is bounded, $J\leq a$,
there are only very few representations to be considered. 
We present them in the following table.
\[\label{eq:SO6desc}
\begin{array}{lc|ccc}
\mbox{Young tab.}&\mbox{Dynkin l.}&\Delta&\Delta-1&\Delta-2
\\\hline
(\Delta,0,0)&[0,\Delta,0]&\irrep{1}&\irrep{4}&\irrep{1}+\irrep{9}
\\
(\Delta-1,1,0)&[1,\Delta-2,1]&&\irrep{4}&\irrep{4}\times \irrep{4}
\\
(\Delta-2,0,0)&[0,\Delta-2,0]&&&\irrep{1}
\\
(\Delta-2,1,+1)&[2,\Delta-3,0]&&&\irrep{3}
\\
(\Delta-2,1,-1)&[0,\Delta-3,2]&&&\irrep{\bar 3}
\\
(\Delta-2,2,0)&[2,\Delta-4,2]&&&\irrep{9}
\end{array}
\]
In the three columns labelled $\Delta-n$ we list
all constituent weights of charge $\Delta-n$ and
the representations of $\grSO(4)$ they form.
Next, we count the number of operators made from $\Delta$ scalars 
in a single trace with charge $J\geq \Delta-2$. 
We note that most of the scalars must be $Z$s, the only 
field without a defect-charge. The scalars $\phi_i$
carry one charge defect and there can only
be two such impurities, $\phi_i, \phi_j$. 
All combinations of indices $i,j$ and positions of defects 
must be considered taking into account the cyclicity of the trace.
The indices take $\irrep{4}$ values and a tensor product of two
can be decomposed into the 
singlet~($\irrep{1}$), antisymmetric ($\irrep{3}+\irrep{\bar{3}}$)
and symmetric-traceless ($\irrep{9}$) part. 
In the trace only the distance between the two impurities matters.
The upper bound for the distance $[(\Delta-2)/2]$ or
$[(\Delta-3)/2]$ for symmetric or antisymmetric combinations
of the impurities.
We obtain the following table
where we list the number of $\grSO(4)$ multiplets
at charge $J=\Delta,\Delta-1,\Delta-2$
\footnote{
There is one additional $\grSO(4)$ singlet due to
the insertion of one double-defect impurity $\bar Z$.}.
\[
\begin{array}{c|ccl}
\grSO(4)&\Delta&\Delta-1&\hfill\Delta-2\hfill
\\\hline
\irrep{1}&1& &1+[\Delta/2]
\\
\irrep{4}& &1&
\\
\irrep{3}& & &\phantom{1+\mathord{}}[(\Delta-1)/2]
\\
\irrep{\bar 3}& & &\phantom{1+\mathord{}}[(\Delta-1)/2]
\\
\irrep{9}& & &1+[(\Delta-2)/2]
\end{array}
\]
By comparing both tables we see that 
the single operator at charge $J=\Delta$ must
transform in the $[0,\Delta,0]$ representation. 
It must be accompanied by a complete multiplet of $\grSO(6)$.
From \eqref{eq:SO6desc} we conclude that also 
the $\irrep{4}$ operators at charge $J=\Delta-1$ 
and $\irrep{1}+\irrep{9}$ operators at charge 
$J=\Delta-2$ belong to the same multiplet.
These are the operators 
$\Tr Z^J, \Tr \phi_i Z^{J-1},\ldots$,
we will come back the explicit form in the next section.
The primary operator at charge $J=\Delta$ 
corresponds to a string-vacuum,
we will thus call its multiplet 
the `vacuum multiplet'.
The $\irrep{4}$ operators at charge $J=\Delta-1$ and
the $\irrep{1}+\irrep{9}$ operators at charge $J=\Delta-2$ 
correspond to strings with one or two excited zero-modes.
Despite that, we will collectively refer to them as vacuum operators
because they belong to one and the same multiplet. 
It is a general feature of the considered multiplets
that they contain not only a fixed number of 
excited non-zero modes (none in this case), but also an arbitrary amount of
excited zero-modes.

All operators at charge $J=\Delta-1$ belong to the
multiplet $[0,\Delta,0]$, hence the multiplet $[1,\Delta-2,1]$
is not realised%
\footnote{
The symmetry of $[1,\Delta-2,1]$
seems to be incompatible with the cyclicity of a single trace.}.
Further operators not contained in the vacuum multiplet 
appear at charge $J=\Delta-2$ and we can identify 
their corresponding $\grSO(6)$ representations.
The following table summarises the multiplets and their
multiplicities up to two charge defects.
\[
\begin{array}{c|cc|l}
\mbox{Dynkin l.}&\grSO(4)&J&\mbox{multiplicity}
\\\hline
{}[0,\Delta,0]&\irrep{1}&\Delta&1
\\
{}[0,\Delta-2,0]&\irrep{1}&\Delta-2&[\Delta/2]
\\
{}[2,\Delta-3,0]&\irrep{3}&\Delta-2&[(\Delta-1)/2]
\\
{}[0,\Delta-3,2]&\irrep{\bar 3}&\Delta-2&[(\Delta-1)/2]
\\
{}[2,\Delta-4,2]&\irrep{9}&\Delta-2&[(\Delta-2)/2]
\end{array}
\]
A central question of this paper is
to what multiplets of supersymmetry the BMN operators belong. 
Due to the multiplicities one can at this point conjecture that
the antisymmetric ($\irrep{3}+\irrep{\bar 3}$)
and symmetric-traceless ($\irrep{9}$) operators
are level $2$ and $4$ supersymmetry descendants of the singlet operators
($\irrep{1}$).
The reason is that two supersymmetry variations 
raise the dimension $\Delta$ by one and the 
multiplicities 
of the shifted multiplets match.

Due to supersymmetry it is not enough to consider only the operators
made from scalars, but 
spinors and gauge fields (in covariant derivatives and
field strengths) have to be taken into account as well. 
The $16$ spinors have dimension $\sfrac{3}{2}$,
half of which ($\psi$) have charge $+\sfrac{1}{2}$ and
half of which ($\bar\psi$) have charge $-\sfrac{1}{2}$.
Therefore, $\psi$ carries one charge defect while 
$\bar\psi$ carries two. 
The $4$ derivatives $\cder_\mu$ carry one charge defect and
the field strength $F_{\mu\nu}$ carries two. 
The following table states the number of all bosonic single-trace operators 
with two charge defects%
\footnote{
At level $8$ the operator with two derivatives at the same
place has been omitted because it can be written as a linear combination
of the other operators via the equations of motion. 
The $\irrep{35}$ at level $4$ is a self-dual four-form of $\grSO(7,1)$
which has not been decomposed to $\grSO(3,1)\times \grSO(4)$ irreps.}.
\[
\begin{array}{c|c|cc|l}
\mbox{level}&\mbox{impurities}&\grSO(3,1)&\grSO(4)&\mbox{multiplicity}
\\\hline\hline
0&\phi\phi,\bar Z\vphantom{\hat{\bar Z}}&\irrep{1}&\irrep{1}&1+[\Delta/2]
\\\hline
2&\psi\psi,F&\irrep{3}+\irrep{\bar 3}\vphantom{\hat{\bar 3}}&\irrep{1}&1+[(\Delta-1)/2]
\\
&\phi\phi&\irrep{1}&\irrep{3}+\irrep{\bar 3}&\phantom{1+}\mathbin{}[(\Delta-1)/2]
\\\hline
4&\phi\phi&\irrep{1}&\irrep{9}&1+[(\Delta-2)/2]
\\
&\psi \psi&\irrep{1}&\irrep{1}&\phantom{1+}\mathbin{}[(\Delta-2)/2]
\\
&\psi \psi&\multicolumn{2}{c|}{\irrep{35}}&\phantom{1+}\mathbin{}[(\Delta-2)/2]
\\
&\cder \cder&\irrep{9}&\irrep{1}&1+[(\Delta-2)/2]
\\\hline
6&\cder \cder&\irrep{3}+\irrep{\bar 3}\vphantom{\hat{\bar 3}}&\irrep{1}&\phantom{1+}\mathbin{}[(\Delta-3)/2]
\\
&\psi\psi&\irrep{1}&\irrep{3}+\irrep{\bar 3}&1+[(\Delta-3)/2]
\\\hline
8&\cder \cder&\irrep{1}&\irrep{1}&1+[(\Delta-4)/2]
\\\hline
2,4,6&\cder \phi&\irrep{4}&\irrep{4}&1+[(\Delta-1)/2]+[(\Delta-2)/2]
\\
&\psi \psi&\irrep{4}&\irrep{4}&1+[(\Delta-3)/2]
\end{array}
\]
From this table we need to subtract components of the vacuum multiplet,
these are indicated as a `$1+$' in the table. 
They are $\grSO(6)$ ($\phi\phi$), supersymmetry ($\psi\psi$),
derivative ($\cder\cder$) or mixed ($\cder\phi$) descendants of
the vacuum operators.
Apart from these we see that the remaining modes
fit nicely into supermultiplets generated by
$8$ fermionic generators. 
This is a strong indication that all these operators
belong to $[\Delta/2]$ supermultiplets of $\superN=4$ SYM 
whose primary operators, $\OpB^{\Delta-2}_n$,
transform in the $[0,\Delta-2,0]$ representation of
$\grSO(6)$ and have naive dimension $\Delta$.
We will show this explicitly in Sec.~\ref{sec:susy},
for the time being we take it as a fact.
These $[\Delta/2]$ supermultiplets 
contain all BMN operators with two defects,
consequently they will be called `BMN multiplets'.
This argument can also be turned around to \emph{define} BMN operators:\smallskip

\emph{
The multiplets of BMN operators 
are the supermultiplets of single-trace operators whose primary operators 
have dimension $\Delta$ and
transform in the ${[0,\Delta-2,0]}$ representation of $\grSO(6)$.
There are $[\Delta/2]$ such multiplets.
}\smallskip

In fact, this is a generalisation of BMN operators which 
previously were defined only in the BMN limit. 
This definition is universal.
Furthermore it describes \emph{only} operators
which are similar to the ones proposed by BMN
\footnote{
It also includes descendant operators with 
more than two impurities, these 
are operators with additional 
zero-modes, but only two 
non-zero modes ($+n,-n$)},
this will be seen in the next section.
In terms of representation theory of 
$\grSU(2\mathord{,}2|4)$ this multiplet belongs to the 
A series of unitary irreducible representations
\cite{Minwalla:1998ka,Arutyunov:2002ff}.
It is at the unitarity bound,
therefore the anomalous dimension is
strictly non-negative. Furthermore, the long supermultiplet 
of $2^{16}\times\dim [0,\Delta-2,0]$ operators
decomposes into a sum of 
shorter multiplets at vanishing coupling constant $\gym=0$
\cite{Dobrev:1985vh}.
In contrast, the vacuum multiplet is a short supermultiplet 
of the C series. As such it is protected, its
scaling dimension cannot be modified by quantum corrections.

%%%%%%%%%%%%%%%%%%%%%%%%%%%%%%%%%%%%%%%%%%%%%%%%%%%%%%%%%%%%%%%%%%%%%%%%%%%%%%%%%
\section{BMN Operators at finite charge}\label{sec:form}

In the last section we have found that 
the single-trace operators with no more than two
defects form one short supermultiplet
$[0,\Delta,0]$ and $[\Delta/2]$ long supermultiplets
$[0,\Delta-2,0]$. 
We need to transform this abstract finding into an explicit form 
for the operators. 

The unique operator without a charge defect is 
$\Tr Z^\Delta$. It is the primary operator of the
supermultiplet $[0,\Delta,0]$ and its $\grSO(6)$ descendants 
with one and two defects are
\footnote{
The superscript $J$ denotes the $\grSO(2)$ charge of the primary 
operator of the supermultiplet. A number in square
brackets denotes an $\grSO(6)$ descendant and a
number in round brackets denotes a supersymmetry 
descendant. The subscripts are mode numbers as well as 
$\grSO(3,1)$ spacetime and 
$\grSO(4)$ internal indices. Round brackets correspond to
symmetric-traceless combinations and 
square brackets to antisymmetric combinations. 
}
\<\label{eq:ZeroDesc}
\OpV^J\eq\frac{N_0^{-J}}{\sqrt{J}}\Tr Z^J
\nln
\OpV^{J,[1]}_{i}\eq N_0^{-J}\Tr \phi_i Z^{J-1}
\nln
\OpV^{J,[2]}\eq \frac{N_0^{-J}}{2\sqrt{2}\,\sqrt{J+1}}
\lrbrk{\sum_{p=0}^{J-2}\Tr \phi_{i} Z^p \phi_{i}Z^{J-2-p}
-4\Tr \bar Z Z^{J-1}}
\nln
\OpV^{J,[2]}_{(ij)}\eq \frac{N_0^{-J}}{\sqrt{2}\,\sqrt{J-1}}
\sum_{p=0}^{J-2}\Tr \phi_{(i} Z^p \phi_{j)}Z^{J-2-p}
\>
with a normalisation constant
\[
N_0=\sqrt{\frac{\gym^2 N}{8\pi^2}}.
\]

From the discussion of the last section we know that
there are $[\Delta/2]$ $\grSO(4)$ singlet operators 
with two scalar impurities 
but we do not know their explicit form.
All of these operators have common quantum numbers 
and mix with each other. 
The explicit form of the operators can be 
determined by requiring two things.
First, the two-point function of such operators
should be canonically normalised at tree level and,
second, the operators should have definite scaling dimension.
From a one-loop, planar calculation at finite $J$ 
we find the forms of the BMN operators with two
scalar impurities
\footnote{
We find $J+3$ in 
the denominator of the cosine of the
singlet operator. This is (hopefully)
the upper bound in the sequence of 
previously suggested denominators $J$ 
\cite{Berenstein:2002jq}, 
$J+1$ \cite{Bianchi:2002rw} and $J+2$ \cite{Parnachev:2002kk}.}
\footnote{
These operators are made out of scalars
only. In fact, one should also consider 
spinor and vector operators belonging to the
same $SO(4)$ representations.
We expect that mixing between these operators becomes relevant 
at higher loops, it is irrelevant in the present investigation.}
\<\label{eq:BMNScalar}
\OpB^J_n\eq
\frac{N_0^{-J-2}}{\sqrt{J+3}}
\bigg[
\half\sum_{p=0}^J\cos\frac{\pi n(2p+3)}{J+3}
\Tr \phi_i Z^p \phi_i Z^{J-p}
-2\cos\frac{\pi n}{J+3}\Tr \bar Z Z^{J+1}
\bigg],
\nln
\OpB^{J-1,(1)}_{[ij],n}\eq
\frac{N_0^{-J-2}}{\sqrt{J+2}}
\sum_{p=0}^{J}
i\sin\frac{\pi n(2p+2)}{J+2}\Tr \phi_{[i}Z^p\phi_{j]} Z^{J-p},
\nln
\OpB^{J-2,(2)}_{(ij),n}\eq
\frac{N_0^{-J-2}}{\sqrt{J+1}}
\sum_{p=0}^{J}
\cos\frac{\pi n(2p+1)}{J+1}\Tr \phi_{(i}Z^p\phi_{j)} Z^{J-p}.
\>
They correspond to the operators
$\mathcal{O}^J_{\irrep{1},n}$, $\mathcal{O}^J_{[ij],n}$
and $\mathcal{O}^J_{(ij),n}$ of \cite{Beisert:2002bb},
the modified notation is more suitable for the 
multiplet structure of the operators.
Interestingly, the one-loop computation at finite $J$ requires 
different phase factors for the three flavours of operators to
acquire complete diagonalisation. 
This is not an issue in the BMN limit, where
the precise form of the phase factors becomes irrelevant.
We note that our phase factors do not agree with the phase factors used in 
the $1/J$ expansion of \cite{Parnachev:2002kk}.
This is not a contradiction, however, 
as the operators in \cite{Parnachev:2002kk}
are explicitly not diagonalised (bare), whereas 
our claim is that the operators
\eqref{eq:BMNScalar} are
the one-loop planar approximation to the full operators.
We will come back to this issue in the conclusions. 

The mode numbers $n$ obey the symmetries
(at this point we shift the dimension of the antisymmetric 
and symmetric-traceless operators by one and two, respectively)
\[\arraycolsep0pt\begin{array}{lclcllcl}
\OpB^J_n\eq+\OpB^J_{-n}\eq-\OpB^J_{J+3+n},&
\OpB^{J}_{(J+3)/2}\eq0,
\\[6pt]
\OpB^{J,(1)}_{[ij],n}\eq -\OpB^{J,(1)}_{[ij],-n}\eq-\OpB^{J,(1)}_{[ij],J+3+n},&
\OpB^{J,(1)}_{[ij],(J+3)/2}\eq0,
\\[6pt]
\OpB^{J,(2)}_{(ij),n}\eq +\OpB^{J,(2)}_{(ij),-n}\eq-\OpB^{J,(2)}_{(ij),J+3+n},\qquad&
\OpB^{J,(2)}_{(ij),(J+3)/2}\eq0
\end{array}\]
and the zero-modes vanish or belong to the vacuum multiplet
\[
\OpB^J_0=\sqrt{2}\,\OpV^{J+2,[2]},\qquad
\OpB^{J,(1)}_{[ij],0}=0,\qquad
\OpB^{J,(2)}_{(ij),0}=\sqrt{2}\,\OpV^{J+4,[2]}_{(ij)}.
\]
To avoid complications concerning linearly dependent operators
we constrain the mode number 
of the operator to the first 
Brillouin zone $0<n<(J+3)/2$.
The number of modes, $[(J+2)/2]$,
matches the number of different operators
as discussed in the last section.
Hence the vacuum and BMN operators form a complete basis
for the space of all operators with two scalar impurities.
It is interesting to see that 
all operators participate in one and the same mode decomposition.
There is no exceptional operator at two defects,
the operator $\Tr \bar Z Z^{J+1}$ mixes with the 
other operators. This means that in the BMN limit
essentially all operators behave similarly,
e.g. have finite anomalous dimensions.

There is a nice alternative 
way to write the singlet BMN operator 
\[
\OpB^J_n=
\frac{N_0^{-J-2}}{2\sqrt{J+3}}
\sum_{p=0}^J\sum_{m=1}^6\cos\frac{\pi n(2p+3)}{J+3}
\Tr \Phi_m Z^p \Phi_m Z^{J-p}
\]
involving a sum over all six scalar fields $\Phi_m$.
Intriguingly, it fails for the zero-mode, $n=0$, 
which belongs to a different multiplet.
We have calculated the two-point functions of these operators 
up to one loop and on the sphere using the effective 
vertices of \cite{Beisert:2002bb} and found 
\<\label{eq:BMNScalar2pt}
\bigvev{\bar\OpB^J_m(x)\,\OpB^K_n(y)}\eq
\frac{\delta_{JK}\,\delta_{mn}}{(x-y)^{2\Delta_n^J}},
\nln
\bigvev{\bar\OpB^{J,(1)}_{[ij],m}(x)\,\OpB^{K,(1)}_{[kl],n}(y)}\eq
\frac{\delta_{JK}\,\delta_{mn}\,\delta_{i[k}\delta_{l]j}}{(x-y)^{2\Delta_n^{J}+2}},
\nln
\bigvev{\bar\OpB^{J,(2)}_{(ij),m}(x)\,\OpB^{K,(2)}_{(kl),n}(y)}\eq
\frac{\delta_{JK}\,\delta_{mn}\,\delta_{i(k}\delta_{l)j}}{(x-y)^{2\Delta_n^{J}+4}}.
\>
The planar one-loop scaling dimension is given by 
\<\label{eq:BMNDimSphere}
\Delta^J_n=J+2+\frac{\gym^2 N}{8\pi^2}\,8\sin^2\frac{\pi n}{J+3}.
\>
Interestingly, we notice that the anomalous piece
is strictly bounded from above by $\gym^2 N/\pi^2$
(at this order in perturbation theory).

We see that the singlet, $\OpB^J_n$, antisymmetric,
$\OpB^{J,(1)}_{[ij],n}$, and
symmetric-traceless,
$\OpB^{J,(2)}_{(ij),n}$, operators 
of naive dimension
$J+2$, $J+3$ and $J+4$ have the same
anomalous dimension $\delta\Delta^J_n$.
This supports our previous conjecture
that these operators belong to the same
supermultiplet
which will be shown explicitly by supersymmetry variations in the next section
\footnote{\label{fn:dearmrreferee}In fact, these operators
are just the leading order (in $\gym$ and $1/N$) approximations
to the exact operators which are exactly related by supersymmetry.
We will show that the approximate operators,
as given in \eqref{eq:BMNScalar}, are related by 
supersymmetry. As being related by supersymmetry 
is a discrete statement, it cannot be changed in perturbation theory.
In other words, although the operators receive corrections, 
they will always belong to the same supermultiplet. 
}.
This will prove to all orders 
(in $\gym$, $1/J$ and $1/N$) the 
equality of anomalous dimensions of the operators
\footnote{Due to the $\grSU(2,2|4)$ commutator
$\comm{D}{Q}=\half Q$ the scaling dimensions of 
all members of a supermultiplet differ by half integers.
Consequently, their anomalous pieces are exactly degenerate.
}.
We will also demonstrate that the redefinitions of operators
in $1/N$ \cite{Beisert:2002bb,Constable:2002vq},
see footnote \ref{fn:dearmrreferee}, obey 
certain relations which guarantee that 
the redefined operators are superpartners even at $\order{1/N^2}$.

In the context of the BMN limit one usually 
considers operators of a common naive dimension 
$J+2$. At finite charge the 
singlet, antisymmetric and symmetric-traceless operators 
have non-degenerate anomalous dimensions
$\delta\Delta_n^J$, 
$\delta\Delta_n^{J-1}$ and 
$\delta\Delta_n^{J-2}$. 
If the BMN limit of $\delta\Delta_n^J$ exists 
to all orders in perturbation theory,
$J$ can appear only in the combinations
$\lambda'$ and $\gtwo$. 
Substituting $J\to J-1,J-2$ only gives rise 
to $\order{1/J}$ corrections
which are irrelevant in the strict BMN limit. 
Therefore all flavours of BMN operators 
have degenerate anomalous dimensions in 
the BMN limit.
From \eqref{eq:BMNScalar2pt} and \eqref{eq:BMNDimSphere} we see 
this explicitly,
namely the anomalous dimension of all operators is $\lambda' n^2$.

Taking a closer look at operators with small $J$ we observe that the
$J=0$ BMN supermultiplet 
including the operators
$(\OpB^0_{1},\OpB^{0,(1)}_{[ij],1},\OpB^{0,(2)}_{(ij),1})$
is the Konishi multiplet 
(see e.g. \cite{Andrianopoli:1998ut,Bianchi:2001cm}).
From \eqref{eq:BMNDimSphere} we obtain the
correct anomalous dimension 
\[
\delta\Delta^0_1=\frac{3\gym^2 N}{4\pi^2}.
\]
The dimension $3$ operator has apparently been left 
out in the literature, we find the anomalous dimension $\delta\Delta^1_1=\gym^2 N/2\pi^2$.
The two $J=2$ BMN operators
coincide with the operators 
that have been studied
in \cite{Bianchi:2002rw}, we obtain the same
planar anomalous dimensions
\[
\delta\Delta^2_{1,2}=\frac{\gym^2 N(5\mp \sqrt{5})}{8\pi^2}.
\]
This is a new connection between the BMN limit and 
operators of a small dimension.
As we shall see later, the
BMN operators with large charge $J$ behave much like
big brothers of the Konishi operator in other respects as well. 
Here, it has proven useful to consider a whole class of operators 
with a fixed number of defects. 
For instance, we have found an expression for a large class of operators and their 
anomalous dimension that interpolates between two regimes.
The number of defects could turn out to be a useful classification 
for operators of $\superN=4$ SYM in general. 
We will address this issue in Sec.~\ref{sec:Concl}.

%%%%%%%%%%%%%%%%%%%%%%%%%%%%%%%%%%%%%%%%%%%%%%%%%%%%%%%%%%%%%%%%%%%%%%%%%%%%%
\section{Supersymmetry}\label{sec:susy}

To show explicitly how the BMN operators are related we
will determine the bosonic supersymmetry 
descendants of the singlet BMN operator.
It is demonstrated that all flavours of two-impurity BMN operators
addressed in Sec.~\ref{sec:BMNmulti} belong to a
single long supermultiplet of $\superN=4$ SYM. 
This puts the previously discovered degeneracy of
anomalous dimension on firm ground.

We will start by short review of 
superspace gauge theory. This will be required
as we would like to take supersymmetry descendants by acting with
fermionic derivatives. As opposed to 
(global) supersymmetry it enables us to 
perform variations on a single 
operator in a correlator and not all
operators together. 
Once a correlation function with full
dependence on the fermionic coordinates is known 
it enables us to derive correlators for all
descendant operators. 
The form of some correlation functions will be inferred from
superconformal symmetry in Sec.~\ref{sec:corr}.

%%%%%%%%%%%%%%%%%%%%%%%%%%%%%%%%%%%%%%%%%%%%%%%%%%%%%%%%%%%%%%%%%%%%%%%%%%%%%
\subsection{Review of Superspace Gauge Theory in $D=9+1$}
\label{sec:supergauge}

We start by reviewing $\superN=1$ gauge theory in $D=9+1$ superspace,
cf. \cite{Gates:1987is,Sohnius:1978wk,Gates:1980jv,Gates:1980wg}.
These papers state that the constraints that have
to be imposed on this theory force the gauge fields on shell.
Therefore, the constraints cannot be solved 
and the theory is not suited well for quantisation.
We are not trying to accomplish this here, all
computations of correlation functions were done 
in non-supersymmetric component language as in 
\cite{Kristjansen:2002bb,Beisert:2002bb}.
Nevertheless, it is a nice framework which allows
one to write down operators and all their 
supersymmetry descendants in a compact way. 
We prefer a ten-dimensional notation over
a four-dimensional one, essentially because it does not have a
distinction of chiralities, resulting in more unified expressions.
In App.~\ref{sec:spinors} we present our notation 
and a collection of useful identities. 

Superspace is parametrised by the
$\irrep{10}$ real bosonic coordinates $X^M$ and the $\irrep{16}$ 
real fermionic coordinates $\Theta^A$. 
Translations on this space are generated by the operators
\[
iP_M=\partial_M,\qquad
iQ_A=\partial_A-\Sigma^M_{AB}\Theta^B\partial_M.
\]
The corresponding supertranslation covariant derivatives 
are
\[
D_M=\partial_M,\qquad
D_A=\partial_A+\Sigma^M_{AB}\Theta^B \partial_M.
\]
The fermionic derivatives satisfy the anticommutation relation
\[
\acomm{D_A}{D_B}=2\Sigma^M_{AB}D_M,
\]
while commutators with a bosonic derivative $D_M$ vanish.

On this space we define a gauge theory with the supercovariant
derivatives
\[
\cder_M=D_M+iA_M,\qquad\cder_A=D_A+iA_A.
\]
Under a gauge transformation $g(X,\Theta)$ the gauge fields transform
canonically according to $A\mapsto gAg^{-1}+i Dg \,g^{-1}$. 
The covariant field strengths of the gauge field are
\<
\acomm{\cder_A}{\cder_B}\eq 2\Sigma^M_{AB}\cder_M+iF_{AB},\nln
\comm{\cder_A}{\cder_M}\eq iF_{AM},\nln
\comm{\cder_M}{\cder_N}\eq iF_{MN}.
\>
We can now impose a constraint on the gauge field, namely that the 
field strength $F_{AB}$ vanishes
\<\label{eq:constraint}
F_{AB}=0.
\>
This field strength can be decomposed 
into two $\grSO(9,1)$ irreps, $\irrep{10}$ and $\irrep{126}$.
The vanishing of the $\irrep{10}$ part determines the bosonic gauge field
$A_M$ in terms of the fermionic one. The $\irrep{126}$ part
has much more drastic consequences as it 
not only reduces the number of independent components, but also
implies equations of motion for the gauge field.
Before stating these, we 
present two important consequences of the constraint
\<\label{eq:susyrel}
\comm{\cder_A}{\cder_M}\eq iF_{AM}=i\Sigma_{M,AB}\Psi^B,
\nln
\acomm{\cder_A}{\Psi^B}\eq
\sfrac{i}{2}\tilde\Sigma^{MN,B}{}_{A}[\cder_M,\cder_N]
=-\sfrac{1}{2}\tilde\Sigma^{MN,B}{}_{A}F_{MN}.
\>
The first shows that the $\irrep{144}$ part of the 
field strength $F_{AM}$ is zero, it 
can be proved by using the Jacobi identity and 
inserting the constraint.
The second one can be proved by projecting on
the $\irrep{1}$, $\irrep{45}$, $\irrep{210}$ parts and using 
the Jacobi identity and constraint. 

It has been shown that the only independent components of the
gauge field are the $\Theta=0$ components 
$A^0_M=\lreval{A_M}_0$ and $\Psi^0=\lreval{\Psi}_0$.
All other components can be gauged away or 
are bosonic derivatives of the fundamental fields. 
%For example, the low-lying components of the gauge fields and field
%strengths are 
%%
%\<
%A\eq
%\Sigma^M\Theta A^0_M
%-\sfrac{1}{36}\Sigma_{MNP}\Psi^0\,\Theta^\trans\Sigma^{MNP}\Theta
%-\sfrac{1}{8}\Sigma_P \Theta\,\Theta^\trans\Sigma^{MNP}\Theta\,  F^0_{MN}
%+\ldots
%\nln
%A_M\eq 
%A^0_M
%+\Theta^\trans\Sigma_M\Psi^0
%-\sfrac{1}{4}\Theta^\trans\Sigma_{MNP}\Theta\,F^{0NP}
%+\ldots
%\nln
%\Psi\eq \Psi^0-\half \Sigma^{MN}\Theta\, F^0_{MN}+\ldots
%\>
%%
The equations of motion which follow from 
\eqref{eq:constraint} 
in much the same way as 
\eqref{eq:susyrel}
are
\<
\comm{\cder_{N}}{F^{NM}}
\eq -\sfrac{i}{2}\Sigma^M_{AB}\acomm{\Psi^A}{\Psi^B}
\nln
\Sigma^M_{AB}\comm{\cder_M}{\Psi^B}\eq 0,
\>
their $\Theta=0$ part forces the fundamental fields $A^0_M$
and $\Psi^0$ on shell.

Using the relations \eqref{eq:susyrel}
one can show that the combination
\footnote{
Note: $\Lagr$ is a descendant of $\Tr \Phi_{(m}\Phi_{n)}$.}
\[
\Lagr=\Tr \bigbrk{\quarter F^{MN} F_{MN}+ 
\half \Psi^A\Sigma^M_{AB}\comm{\cder_M}{\Psi^B}}
\]
has the property that the fermionic derivative is
a total bosonic derivative,
$\comm{D_A}{\Lagr}=\comm{\partial_M}{B^M_A}$
and the action 
\[
S=\frac{2}{\gym^2}\int d^D X \, \bigeval{\Lagr}_0
\]
is thus supersymmetric.

%%%%%%%%%%%%%%%%%%%%%%%%%%%%%%%%%%%%%%%%%%%%%%%%%%%%%%%%%%%%%%%%%%%%%%%%%%%%%
\subsection{A Harmonic Superspace}

We introduce a harmonic superspace which is adapted to
the treatment of BMN operators. It consists of
the usual superspace coordinates plus 
additional bosonic coordinates parametrising the coset
$\grSO(6)/\grSO(4)$. It enables us to work with the
$\grSO(4)\times\grSO(2)$ split of BMN operators 
while keeping $\grSO(6)$ invariance.

First, we reduce the $D=9+1$ superspace to four dimensions,
$X^M=(x^\mu,0)$, to obtain $\superN=4$ super Yang-Mills theory in 
$D=3+1$ dimensions. The vector indices $M$ split up into
$\mu=0,\ldots,3$ and $m=1,\ldots,6$, while we 
keep the spinor indices $A$
(most of the time they will be suppressed in matrix notation). 
The six gauge fields corresponding to the reduced
coordinates, $A_m$, become the six scalar fields 
$\Phi_m=A_m$ of $\superN=4$ SYM.
The field strengths with internal indices are
$F_{\mu m}=\cder_\mu \Phi_m$ and
$F_{mn}=i[\Phi_m,\Phi_n]$.

In addition to the superspace coordinates $z=(x^\mu,\Theta^A)$.
We introduce a complex vector $V^m$ in the internal space
with zero square and unit norm,
\[
V^2=0,\qquad |V|^2=1,
\]
for example $V=(0,0,0,0,\sfrac{1}{\sqrt{2}},\sfrac{i}{\sqrt{2}})$
\footnote{One could also set
$V=\bigbrk{\tau_1,\tau_2,\tau_3,\tau_4,
\sfrac{1}{\sqrt{2}}(1-\half\tau^2),\sfrac{i}{\sqrt{2}}(1+\half\tau^2)}$
with four parameters $\tau_k$. 
One can then define
$\grSO(6)$ primary fields in much the same way as
conformal primary fields
\cite{Arutyunov:2002xd}. 
The conjugate $\bar V$ is obtained by $\tau$ inversion as
$V(-1/\tau)=-\bar V(\tau)/\tau^2$ and $\grSO(6)$ descendants
are obtained by acting with $\partial/\partial \tau_k$.
The $\grSO(6)$ correlator of unit dimension is 
$V_m(\tau) V^m(\tau')=-\half(\tau-\tau')^2$.}.
It has the remarkable property that
a product of $J$ vectors $V^m V^n \cdots$
is not only completely symmetric in all
indices, but also completely traceless. 
Therefore, the product transforms in the 
irreducible representation $[0,J,0]$ of 
$\grSO(6)$. 

Assume $A_{mn\ldots}$ is a symmetric-traceless tensor.
Then $A$ can be written as a
holomorphic function in $V$
\[
A(V)=A_{mn\ldots} V^m V^n\cdots
\]
This map is injective, 
the components of the tensor $A$
correspond to the amplitude of
certain harmonics on the space
$\grSO(6)/\grSO(4)$, the space of $V$s.
The vacuum and BMN operators transform 
like $A$ under $\grSO(6)$ and it is very convenient 
to write down these operators as holomorphic 
functions in $V$. 
This keeps $\grSO(6)$ covariance manifest
without making use of indices.
In this sense, the null-vector $V$ is a set of additional bosonic
coordinates of a `harmonic' superspace.
It is adapted to the $\grSO(2)\times\grSO(4)$ split
of BMN operators
in that $\grSO(2)$ acts as a complex phase rotation on $V$
and $\grSO(4)$ leaves $V$ invariant.

With $V$ we can isolate two components,
$a_+$ and $a_-$, of 
an internal space vector $a_m$
\[
a_+=a^-=a_m V^m,\qquad 
a_-=a^+=a_m \bar V^m,\qquad
a_i V^i=a_i \bar V^i=0,
\]
the remaining four components, $a_i$, are labelled by $i=1,\ldots,4$. The 
internal space metric becomes
\[
\delta_{i=j}=\delta_{+-}=\delta_{-+}=1,\quad 
\delta_{i\neq j}=\delta_{i+}=\delta_{i-}=\delta_{++}=\delta_{--}=0.
\]
and the product of two $\Sigma$ matrices in the direction of $V$ vanishes
\[
\Sigma_+\tilde\Sigma_+=\Sigma_{++}+\delta_{++}=0
\]
because two $+$ indices can neither be antisymmetrised nor 
do they have a trace. Using this property it is easily seen that
$\tilde P_+=\half\tilde\Sigma_-\Sigma_+$,
$\tilde P_-=\half\tilde\Sigma_+\Sigma_-$
are orthogonal projection operators which project to 
one half of the spinor space. 

We decompose the $\grSO(6)$ vector of scalar fields into
\[
Z=\Phi_+,\quad
\bar Z=\Phi_-,\quad
\phi_i=\Phi_i,
\]
and the spinor field into
the two spinor subspaces
\[
\psi=\tilde P_+\Psi,\qquad
\bar\psi=\tilde P_-\Psi.
\]

We will be interested in supersymmetry transformations that 
leave the number of charge defects invariant.
This can be done by using the combination 
\[\label{eq:subalgebra}
\delta_\epsilon
=\epsilon^\trans\tilde\Sigma_+\cder
=\epsilon_A\tilde\Sigma^{AB}_+\cder_B,
\]
where $\epsilon_A$ is an arbitrary anticommuting spinor.
The matrix $\tilde\Sigma_+$ projects out the 
parts that would increase the number of defects by one.
This leaves only $8$ independent variations parametrised by $\epsilon$.
It commutes with the combination 
$\Delta-J$ 
\[
\comm{\Delta-J}{\delta}=0
\]
of the generator of dilatations and $\grSO(2)$ rotations, $\Delta$ and $J$.
It thus leaves the number of impurities invariant.
In plane-wave string theory it corresponds to the
supersymmetry generator commonly denoted by $Q^-$.
It is easily seen that the operators $\delta$
commute among each other
\[
\comm{\delta_1}{\delta_2}
=\epsilon_{1,A}\tilde\Sigma^{AB}_+\acomm{\cder_B}{\cder_C}
   \tilde\Sigma^{CD}_+\epsilon_{2,D}
=2\epsilon_1^\trans\tilde\Sigma_+\Sigma^M\tilde\Sigma_+\epsilon_2\,\cder_M
=4i\epsilon_1^\trans\tilde\Sigma_+\epsilon_2\, Z
\]
up to a gauge transformation proportional to the field $Z$.
The algebra generated by the operators in
\eqref{eq:subalgebra} is thus a $\Real^8$
subalgebra of $\grSU(2\mathord{,}2|4)$.
We can therefore act with the same $\delta_\epsilon$
iteratively to obtain all descendants of a gauge 
invariant operator. With the $8$ independent 
fermionic generators
$2^8=256$ descendants are generated unless any
of the variations $\delta^l$, $l\leq 8$, 
annihilates the operator\footnote{
For each of these $256$ operators $256$ independent 
operators can be generated by
the $8$ generators
$\epsilon^\trans\tilde\Sigma_-\cder$
resulting in a long supermultiplet
of $2^{16}$.}.
This, for example, happens for the 
primary vacuum operator. 
Let us write down the variations of the component fields
using the supersymmetry rules \eqref{eq:susyrel}
\<\label{eq:varyfield}
\delta Z\eq 0,
\nln
\delta \psi\eq 
-\tilde\Sigma_-\Sigma_+\tilde\Sigma^\mu \epsilon\, \cder_\mu Z
-i\tilde\Sigma_-\Sigma_+\tilde\Sigma^i \epsilon\, [\phi_i, Z],
\nln
\delta \phi_i\eq
\epsilon^\trans\tilde\Sigma_+\Sigma^i\psi,
\nln
\delta \cder_\mu\eq
i\epsilon^\trans\tilde\Sigma_+\Sigma^\mu\psi,
\nln
\delta \bar\psi\eq 
-i\tilde\Sigma_+\epsilon\,\comm{\bar Z}{Z}
-\sfrac{i}{2}\tilde\Sigma_+\tilde\Sigma^{ij}\epsilon\,\comm{\phi_i}{\phi_j}
-\tilde\Sigma_+\Sigma^\mu\tilde\Sigma^i\epsilon\,\cder_\mu\phi_i
-\half\tilde\Sigma_+\Sigma^{\mu\nu}\epsilon\,F_{\mu\nu},
\nln
\delta \bar Z\eq 2\epsilon^\trans \bar\psi.
\>
%

%%%%%%%%%%%%%%%%%%%%%%%%%%%%%%%%%%%%%%%%%%%%%%%%%%%%%%%%%%%%%%%%%%%%%%%%%%%%%
\subsection{BMN Operators}

In the harmonic superspace notation the primary vacuum
\eqref{eq:ZeroDesc}
and BMN operators \eqref{eq:BMNScalar}
at finite $J$ become
\<
\OpV^J(z,V)\eq
\frac{N_0^{-J}}{\sqrt{J}}\Tr Z^J,
\\\nonumber
\OpB^J_n(z,V)\eq
\frac{N_0^{-J-2}}{\sqrt{J+3}}\bigg[
\half\sum_{p=0}^J\cos\frac{\pi n(2p+3)}{J+3}
\Tr \phi_i Z^p \phi_i Z^{J-p}
-2\cos\frac{\pi n}{J+3}
\Tr \bar Z Z^{J+1}
\bigg]
\>
The first important point to notice is that
the variation of $Z$ vanishes \eqref{eq:varyfield}. 
Thus the variations act only on the impurities,
the string of $Z$s in the operators appears merely as a background.
The immediate consequence is that the vacuum operator
is invariant under this variation
\[
\delta \OpV^J=0.
\]
This is the $\half$ BPS condition.

It is a matter of patience and Fierz identities
to work out the variations of the BMN operators
\<\label{eq:BMNvar}
\delta^2 \OpB^{J}_n\eq 
\sqrt{2}\,\kappa_n^J\,e^{ij} \,\OpB^{J,(1)}_{[ij],n}
+\ldots
\nln
\delta^4\OpB^J_n\eq
-6(\kappa^J_n)^2e^{i\mu}e^{j}{}_{\mu}
\OpB^{J,(2)}_{(ij),n}
+\ldots
\>
This involves the combination of two variation coefficients
\footnote{
The variation coefficients allow one to 
project the most general descendant operator
to one specific operator.
For example, the combination
$(\partial_\epsilon^\trans
\Sigma_{-}\tilde\Sigma_{ij}
\partial_\epsilon)$
projects the general variation 
$\delta^2 \OpB^{J}_n$
to the operator
$\OpB^{J,(2)}_{n,[ij]}$
with the coefficient $e_{ij}$
due to the identity
$(\partial_\epsilon^\trans
\Sigma_{-}\tilde\Sigma_{ij}
\partial_\epsilon)
e^{MN}\sim
\delta_{[i}^{M}\delta_{j]}^{N}$.
}
\[\label{eq:epssqr}
e_{MN}=\epsilon^\trans\tilde\Sigma_{+}\Sigma_{MN}\epsilon,
\]
and a constant
\[
\kappa_n^J=\sqrt{8}N_0\sin\frac{\pi n}{J+3}=
\sqrt{8}\,\sqrt{\frac{\gym^2 N}{8\pi^2}}\,\sin\frac{\pi n}{J+3}\sim
\sqrt{\lambda'}\,n.
\]
The supersymmetry variation \eqref{eq:BMNvar}
is the principal result of this section. 
It shows that the operators 
\<
\OpB^{J,(1)}_{[ij],n}(z,V)\eq
\frac{N_0^{-J-3}}{\sqrt{J+3}}
\sum_{p=0}^{J+1}
i\sin\frac{\pi n(2p+2)}{J+3}\Tr \phi_{[i}Z^p\phi_{j]} Z^{J+1-p},
\nln
\OpB^{J,(2)}_{(ij),n}(z,V)\eq
\frac{N_0^{-J-4}}{\sqrt{J+3}}
\sum_{p=0}^{J+2}
\cos\frac{\pi n(2p+1)}{J+3}\Tr \phi_{(i}Z^p\phi_{j)} Z^{J+2-p},
\>
belong to the same supermultiplet as the primary operator $\OpB^J_n$. 
This proves that they have degenerate anomalous dimension to all orders in 
perturbation theory (see also the discussion at the end of
Sec.~\ref{sec:form}).
Furthermore the opertators coincide with the other 
two operators we obtained by a direct 
calculation in \eqref{eq:BMNScalar}.
We have presented only the descendants with two scalar impurities here,
the complete set of bosonic descendants can 
be found in App.~\ref{sec:BMNDesc}. 
It verifies the claim of Sec.~\ref{sec:BMNmulti} that all a 
BMN multiplet contains all flavours of BMN operators.

The constants appearing in front of the operators were chosen 
such that the two-point function of these operators 
are canonically normalised as in \eqref{eq:BMNScalar2pt}. 
The constant $\kappa^J_n$ is in fact the square root of the
anomalous dimension. Its appearance is related to
a splitting of the long multiplet at $\gym=0$ which happens
at the unitarity bound of the A series of unitary
irreducible representations of $\grSU(2\mathord{,}2|4)$
\cite{Dobrev:1985vh}.

We can also write the above variations in a general form as
\[\label{eq:BMNvargen}
\delta^l \OpB^J_n=\sum_a N^{J,(l/2)}_{a,n} (\epsilon^l)^a \OpB^{J,(l/2)}_{a,n}
\]
Here, $a$ labels the descendant operators at level $l$,
$(\epsilon^l)^a$ is the corresponding variation coefficient
and $N^{J,(l/2)}_{a,n}$ is a normalisation constant.
To be explicit we write down these quantities
for the operators in \eqref{eq:BMNvar}
\[\label{eq:NormEx}
\arraycolsep0pt\begin{array}{rclcrcl}
N^{J,(1)}_{[ij],n}\eq\sqrt{2}\kappa^J_n,&\qquad&
(\epsilon^2)^{[ij]}\eq e^{ij},
\\[5pt]
N^{J,(2)}_{(ij),n}\eq-6\bigbrk{\kappa^J_n}^2,&&
(\epsilon^4)^{(ij)}\eq e^{i\mu}e^{j}{}_{\mu},
\end{array}
\]
%

%%%%%%%%%%%%%%%%%%%%%%%%%%%%%%%%%%%%%%%%%%%%%%%%%%%%%%%%%%%%%%%%%%%%%%%%%%%%%
\subsection{Operator Mixing}

When considering non-planar corrections to correlation
functions one has to take into account that 
operators with different numbers of traces undergo mixing.
The fact that different BMN operators are related to another
by supersymmetry means that
the mixing matrices are also related. 
We will now investigate this relationship.

In \cite{Beisert:2002bb,Constable:2002vq} the issue 
of operator mixing was investigated in the BMN limit up 
to the torus and up to one-loop. 
The resulting expressions 
for the modified operators at $\order{\gtwo}$ are
\footnote{
The expressions found in \cite{Beisert:2002bb} 
had to be (anti)symmetrised in the mode numbers
for the singlet and (anti)symmetric operators to project to
the relevant part of the mixing matrix. }
\<\label{eq:BMNmixing}
\OpB^{\prime\,J}_{n}\eq 
\OpB^J_{n}-\gtwo\sum_{k,r} n^2\,X^{J,r}_{n,k}\,\OpB^{Jr}_{k}\OpV^{J(1-r)}
-\gtwo\sum_r n^2\,Y^{J,r}_{n}\, \half\OpV^{Jr+1,[1]}_{i}\OpV^{J(1-r)+1,[1]}_{i}
\nln
\OpB^{\prime\,J,(1)}_{[ij],n}\eq 
\OpB^J_{n}-\gtwo\sum_{k,r} n\sfrac{k}{r}\,X^{J,r}_{n,k}\,\OpB^{Jr,(1)}_{[ij],k}\OpV^{J(1-r)}
\nln
\OpB^{\prime\,J,(2)}_{(ij),n}\eq 
\OpB^J_{n}-\gtwo\sum_{k,r} \sfrac{k^2}{r^2}\,X^{J,r}_{n,k}\,\OpB^{Jr,(2)}_{(ij),k}\OpV^{J(1-r)}
\nln
X^{J,r}_{n,k}\eq \frac{\sqrt{1-r}\,\sin^2(\pi n  r)}{\sqrt{Jr}\,(n^2-\sfrac{k^2}{r^2})^2},
\qquad
Y^{J,r}_{n}=-\frac{\sin^2(\pi n r)}{\sqrt{J}\,\pi^2 n^2}.
\>
These three expressions are very similar to each other. 
When going from the singlet to the antisymmetric to the symmetric-traceless 
operator the coefficient in front of $\OpB^{Jr}_k\OpV^{J(1-r)}$ 
is multiplied by $k/nr$
and the operator 
$\OpV^{Jr+1,[1]}_{i}\OpB^{J(1-r)+1,[1]}_{j}$ is dropped. 
This pattern is due to supersymmetry.
We apply $\delta^2$ to the right hand side of 
the first equation of 
\eqref{eq:BMNmixing} and
find that the single-trace operator
gets multiplied by the normalisation constant 
$N^{J,(1)}_{[ij],n}$ while the first double-trace 
operator gets multiplied by 
$N^{Jr,(1)}_{[ij],k}$. The second double-trace
operator does not have an $e^{ij}$ descendant and drops out.
The resulting expression is to be compared to the second
line in \eqref{eq:BMNmixing}.
It is easily seen that the expressions match provided that
$\delta^2\OpB^{\prime\,J}_{n}=N^{J,(1)}_{[ij],n}e^{ij}
\OpB^{\prime\,J,(1)}_{[ij],n}+\ldots$, i.e. the same 
variation as for $\OpB^{J}_{n}$,
in \eqref{eq:BMNvar}. 
Effectively this means that only the 
coefficient of the double-trace operator changes
by $N^{Jr,(1)}_{[ij],k}/N^{J,(1)}_{[ij],n}\sim k/nr$.
A similar discussion applies to the third operator in 
\eqref{eq:BMNmixing} and also to the mixing matrix 
of double-trace operators. 
It is interesting to compare the mixing of BMN operators
at large $J$ and small $J$. 
It can be seen by comparing to
\cite{Bianchi:1999ge,Arutyunov:2000ku,Bianchi:2002rw}
that the mixing pattern of single-trace and 
double-trace operators is essentially the same
as in the BMN limit, \eqref{eq:BMNmixing}
\cite{Beisert:2002bb}.

The conclusion is that the variations of the redefined operators 
do not change at $\order{\gtwo}$.
Using the expressions and matrices in \cite{Beisert:2002bb} we find that 
the normalisation constants of the 
variations do change at $\order{\gtwo^2}$,
while the form of the variations 
\eqref{eq:BMNvargen} remains unchanged
due to nontrivial relations between the matrices.
The normalisation constants on the torus are
\<\label{eq:varytorus}
N^{J,(1)}_{[ij],n}\mathrel{}&\sim&\mathrel{} 
\sqrt{2\lambda'}\,n
\lrbrk{1+\gtwo^2\lrbrk{\frac{1}{96\pi^2n^2}+\frac{35}{256\pi^4 n^4}}},
\nln
N^{J,(2)}_{(ij),n}\mathrel{}&\sim&\mathrel{} 
-6 \lambda' n^2
\lrbrk{1+\gtwo^2\lrbrk{\frac{1}{48\pi^2n^2}+\frac{35}{128\pi^4 n^4}}}.
\>
As we shall see later this modification is related
to a modification of the anomalous dimension on the torus. 
In the following we shall consider only the full operators 
after a complete normalisation and diagonalisation. 
Certainly, this can be done in a perturbative fashion 
and the operators as defined in 
\eqref{eq:BMNScalar} are just their lowest-order 
approximations. 

%%%%%%%%%%%%%%%%%%%%%%%%%%%%%%%%%%%%%%%%%%%%%%%%%%%%%%%%%%%%%%%%%%%%%%%%%%%%%
\section{Correlation functions}\label{sec:corr}

In this section we will consider correlation functions 
involving BMN operators and restrictions on their form 
imposed by superconformal symmetry.
The aim is to obtain the form of correlation functions in superspace
and apply it to relate correlators of BMN operators 
which have been calculated \cite{Beisert:2002bb,Constable:2002vq}. 
We will start by reviewing some results of $\superN=4$ superconformal
symmetry and later apply them to correlators of BMN operators. 

%%%%%%%%%%%%%%%%%%%%%%%%%%%%%%%%%%%%%%%%%%%%%%%%%%%%%%%%%%%%%%%%%%%%%%%%%%%%%
\subsection{Review of Superconformal Symmetry}

Conformal symmetry in four spacetime dimensions has been addressed 
in the works
\cite{Todorov:1978rf,Erdmenger:1997yc,Osborn:1998qu,Park:1999pd} with 
different numbers of supersymmetries.
We will start by reviewing some results 
of \cite{Park:1999pd} in the notation of that paper.

In this context it is useful to consider $\superN=4$ superspace as 
a coset space of the supergroup $\grSU(2\mathord{,}2|4)$ over the subgroup 
generated by spacetime rotations, internal rotations, 
dilatations and superconformal boosts.
Due to this coset space construction it is natural to
believe that $\superN=4$ superspace is not flat, i.e.
tangent vectors at different points in superspace
cannot be compared directly. 
For correlators of operators with a
tensor structure, however, this is exactly what one 
needs to do. The tensor indices at one
point in superspace need to be saturated by tensor 
indices at some other point. 

In \cite{Park:1999pd} expressions 
for the connection of $\grSU(4)$ internal 
and $\grSL(2,\Comp)$ spacetime spinors
were given
\[\label{eq:SU4conn}
\hat v^a{}_b(z_{12})=\lrbrk{\frac{x_{\bar 21}^2}{x_{\bar 12}^2}}^{1/4}v^a{}_b(z_{12})
\in \grSU(4),\qquad
\hat{\tilde x}_{\bar 12}=\frac{\tilde x_{\bar 12}}{|x_{\bar 12}|}
\in \grSL(2,\Comp)
\]
where
\[\label{eq:vconn}
v^a{}_b(z_{12})=\delta^a{}_b+4i\theta^a_{12}\tilde x_{\bar 12}^{-1}\bar\theta_{12b}.
\]
Using the invariant tensors $\sigma^m_{ab}$ an 
$\grSO(6)$ vector index can be transformed into two antisymmetric 
$\grSU(4)$ spinor indices. 
In this way internal vectors at two different 
points in superspace can be related with the connection 
\[
J_{12,mn}=
\quarter\lrbrk{\frac{x_{\bar 21}^2}{x_{\bar 12}^2}}^{1/2}
\sigma_{m,ab}\, v^a{}_c(z_{12}) v^b{}_d(z_{12})\,\sigma_{n}^{cd} 
\in \grSO(6).
\]
Due to the self-duality of the $\irrep{6}$ representation of
$\grSU(4)$ the internal vector connection can also be written 
in a different fashion
\[
J_{12,mn}=
\quarter\lrbrk{\frac{x_{\bar 12}^2}{x_{\bar 21}^2}}^{1/2}
\sigma_{n,ab}\, v^a{}_c(z_{21}) v^b{}_d(z_{21})\,\sigma_{m}^{cd} 
\in \grSO(6).
\]
Equivalently, for spacetime vectors there is the connection
\[
J_{12,\mu\nu}=\half\Tr \sigma_\mu\hat{\tilde{x}}_{\bar 12}\sigma_\nu\hat{\tilde{x}}_{\bar 21}
=\half\Tr \tilde\sigma_\mu\hat{x}_{\bar 21}\tilde\sigma_\nu\hat{x}_{\bar 12}.
\]
The $\theta=\bar\theta=0$ components of $P_{12}$, $J_{12,mn}$ and $J_{12,\mu\nu}$ 
are given by 
\[\label{eq:proptop}
\bigeval{P_{12}}_{0}=\frac{1}{(x_1-x_2)^2},\qquad
\bigeval{J_{12,mn}}_{0}=\delta_{mn},\quad
\bigeval{J_{12,\mu\nu}}_{0}=\eta_{\mu\nu}-2\,\frac{(x_1-x_2)_\mu (x_1-x_2)_\nu}{(x_1-x_2)^2}.
\]

In correlation functions internal space vectors are usually due to
the field $\Phi_m$ of unit dimension. Therefore one is tempted 
to combine $J_{12,mn}$ with the superconformal scalar correlator 
of unit dimension
\[
P_{12}=\frac{1}{(x_{\bar 12}^2 x_{\bar 21}^2)^{1/2}}
\]
to 
\[\label{eq:propconn}
K_{12,mn}
=P_{12}J_{12,mn}
=
\frac{\sigma_{m,ab}\, v^a{}_c(z_{12}) v^b{}_d(z_{12})\,\sigma_{n}^{cd} }{4x_{\bar 12}^2}
=
\frac{\sigma_{n,ab}\, v^a{}_c(z_{21}) v^b{}_d(z_{21})\,\sigma_{m}^{cd} }{4x_{\bar 21}^2}.
\]
This is the unique two-point function of unit conformal 
dimension at points $z_1$ and $z_2$ 
which relates an internal vector index $m$ at point $z_1$ with
the index $n$ at point $z_2$ in a superconformally covariant way. 
An interesting feature of the first form of the function $K_{12,mn}$ is 
that it depends only on the coordinates
$x_{\bar 12}$ and $\bar\theta_{12}$, which are
anti-chiral at point $z_1$. The only exception is
a chiral $\theta_{12}^a$ in $v^a{}_b(z_{12})$, \eqref{eq:vconn},
a chiral derivative $D_1$ with respect to $z_1$ 
will therefore act only on this. 
Now assume that we act with the combination
$V^p\sigma_p^{ba} D_{1,a\alpha}$ on 
$V^m K_{12,mn}$ where $V$ is a (complex)
null-vector of $\grSO(6)$.
We then find that
the result is proportional to 
$V^p\sigma_p^{ba} V^m \sigma_{m,ac}=V^2 \delta^b_c=0$,
i.e.
\[
(V^p\sigma_p^{ba} D_{1,a\alpha}) (V^m K_{12,mn})=0 \quad\mbox{for }V^2=0.
\]
A similar argument holds for the second form of $K$
in \eqref{eq:propconn}
and anti-chiral derivatives
\[
(V^p\sigma_{p,ba} \bar D^a_{1,\alpha}) (V^m K_{12,mn})=0 \quad\mbox{for }V^2=0.
\]
This remarkable property shows that
$V^m K_{12,mn}$ is invariant under half the supersymmetry.
We will make excessive use of it in the context of
the vacuum operators which are $\half$ BPS.

Two-point functions of superconformal (quasi)primary operators
are uniquely determined by the representation and
conformal dimension of the operator. 
To construct a two-point function 
the representations of the spacetime and 
internal group have to be transformed into a tensor product 
of spinor representations, The spinor indices are then
contracted by the $\grSL(2,\Comp)$ or $\grSU(4)$ 
connections \eqref{eq:SU4conn}. 
This is to be multiplied by $P_{12}^\Delta$, where
$\Delta$ is the scaling dimension of the operator. 
We are dealing only with operators 
in trivial spacetime representations
and tensor product representations of $\grSO(6)$ vectors. 
Therefore, the vector indices can be saturated by 
powers of $K_{12,mn}$ and the remaining scaling dimension by
powers of $P_{12}$.

Superconformal symmetry does not determine three-point functions 
of superconformal primary operators uniquely. 
There are, however, 
some principles which constrain their form. 
First of all the conformal dimension at each of the
three points must match the dimension of the operator at
that point. This can always be achieved by 
products of scalar propagators $P_{12}$, $P_{23}$ and $P_{31}$.
Secondly, the indices of the operators need
to be connected to each other in a covariant way. 
The most general way to accomplish this is to
parallel transport the indices of two operators 
to the point of the third one. 
At that point the indices can be contracted with each 
other or with a covariant 
superspace tangent vector, $\mathbb{Z}$. 
We will, however, proceed differently. 
We construct three-point functions 
as products of two-point functions between all three pairs
of points. Examples of this are given in \cite{Howe:1998zi,Eden:1999gh}. 
It is not clear to the author whether this is 
as general as the above construction,
nevertheless, there is reasonable freedom as we shall see below.
We will use the above construction rules as a guideline 
to derive possible three-point functions of the vacuum and
BMN operators. This will then turn out to reproduce explicitly 
computed correlators exactly.

An interesting feature of three-point functions is
that two connections can be joined to obtain 
a different connection from the direct one, e.g.
\[
v^a{}_b(z_{12})v^b{}_c(z_{23})\neq v^a{}_c(z_{13}).
\]
This is not possible for two-point functions, 
as there are only two points and for $z_3=z_1$ this combination is 
the unit matrix.
One can furthermore construct longer chains of connections
which can either end in indices of the operators or 
be closed. A closed chain turns out to be a combination of the
two $\superN=4$ superconformal three-point invariants.

%%%%%%%%%%%%%%%%%%%%%%%%%%%%%%%%%%%%%%%%%%%%%%%%%%%%%%%%%%%%%%%%%%%%%%%%%%%%%
\subsection{Two-Point Functions}

With these results we consider two-point functions 
of BMN operators in the reduced ten-dimensional 
notation of the last section (see also App.~\ref{sec:spinors}). 
In this notation the supertranslation invariant 
interval $z_{12}=(x_{12}^\mu,\Theta_{12}^A)$, 
$(Q_1+Q_2)z_{12}=0$, is
\[
x_{12}^\mu=x_1^\mu-x_2^\mu-\Theta_1^\trans\Sigma^\mu\Theta_2,\qquad
\Theta_{12}^A=\Theta_1^A-\Theta_2^A.
\]
The combination $x_{\bar 12}$ of \cite{Park:1999pd} corresponds to 
\[
y_{12}^\mu=x_{12}^\mu
-\sfrac{i}{12}\eps^{\mu\nu\rho\sigma}\Theta_{12}^\trans\Sigma_{\nu\rho\sigma}\Theta_{12}
=x_{12}^\mu+\sfrac{i}{2}\Theta_{12}^\trans \Sigma_{(4)}\Sigma^\mu\Theta_{12},
\]
in our language,
where $\Sigma_{(4)}=\sfrac{1}{24}\eps^{\mu\nu\rho\sigma}\Sigma_{\mu\nu\rho\sigma}$.
Its conjugate 
$\bar y_{12}=x_{12}^\mu-\sfrac{i}{2}\Theta_{12}^\trans \Sigma_{(4)}\Sigma^\mu\Theta_{12}$ 
corresponds to $-x_{\bar 21}$.

With these we can construct the scalar propagator 
\[\label{eq:scaprop}
P_{12}=\frac{1}{(y_{12}^2\bar y_{12}^2)^{1/2}}
\]
of unit dimension at points $z_1$ and $z_2$.
The propagator of unit dimension correlating two $\grSO(6)$ vectors
is some function $K_{12,mn}$. It has the property that 
the variations
\[
\delta_1=V_1^m \epsilon_{1}^\trans\tilde\Sigma_m D_{1},\quad
\delta_2=V_2^m \epsilon_{2}^\trans\tilde\Sigma_m D_{2}
\]
annihilate the combinations
\[\label{eq:KhalfBPS}
\delta_1 \bigbrk{V_1^m K_{12,mn}}=0, \qquad
\delta_2 \bigbrk{V_2^n K_{12,mn}}=0.
\]
This is all we need to know about it.
Superconformally covariant two-point functions of 
operators involving only internal vector indices can be
constructed from these two building blocks.

Superconformal symmetry fixes two-point functions 
of superconformal (quasi)primary operators uniquely.
The operators $\OpV^J$ and $\OpB^J_n$ are 
superconformal primaries, essentially 
because they cannot be a descendant of any operator,
see the discussion in Sec.~\ref{sec:BMNmulti}.
We construct a conformally covariant 
two-point function for the vacuum operator $\OpV^J$
from the above building blocks.
Its $\grSO(6)$ representation is $[0,J,0]$, 
the symmetric-traceless tensor product of $J$ vectors. 
The scaling dimension $\Delta=J$ is protected from 
acquiring quantum corrections. 
Consequently, the unique two-point function 
is
\[\label{eq:KBPS}
\bigvev{\OpV^J(z_1,V_1)\,\OpV^J(z_2,V_2)}=K_{12}^J
\]
with
\[
K_{12}=V_1^m V_2^n K_{12,mn}.
\]
Here we use independent null-vectors $V_1$, $V_2$ to be able
to project each operator to an arbitrary component of
the $\grSO(6)$ multiplet. To compare to
explicit computations we set $V_1=\bar V_2$,
because $\OpV(z,\bar V)=\OpV(z,V)^\conj$.
Eq. \eqref{eq:KhalfBPS} implies that
\[
\delta_1 \bigvev{\OpV^J(z_1,V_1)\,\OpV^J(z_2,V_2)}=
\delta_2 \bigvev{\OpV^J(z_1,V_1)\,\OpV^J(z_2,V_2)}=0.
\]
reflecting the $\half$ BPS property $\delta \OpV^J=0$.

In the case of the BMN operators $\OpB^J_n$ there are
$J$ indices from the
representation $[0,J,0]$ to be connected
in just the same way as for the vacuum operators. 
The conformal dimension, however,
is not saturated by this, we need to multiply
by powers of $P_{12}$ to match it.
The correlator is thus
\[
\bigvev{\OpB^J_n(z_1,V_1)\,\OpB^J_n(z_2,V_2)}=
K_{12}^J \,P_{12}^{\Delta^J_n-J}
\]
Again the 
combination $K_{12}$ is invariant 
under $\delta_1, \delta_2$ and
just like the string of $Z$s in the 
definition of the operators 
\eqref{eq:BMNScalar} it may be viewed as a background.
What remains is the propagator 
of a Konishi-field (with modified 
anomalous dimension). The BMN operators 
could therefore be viewed as
Konishi operators in a background 
that provides large dimension and charge.

We now perform a direct calculation of some descendant 
correlators. 
To this end we need to consider only the $P_{12}$ part
of the two-point function as $K_{12}$ is invariant under the
variations. 
First we work out the variation on $y_{12}$
\[
\delta_1 y_{12}^\mu=2\epsilon_1^\trans\tilde\Sigma_{+}P\Sigma^\mu\Theta_{12},
\]
where $P=\half+\sfrac{i}{2}\Sigma_{(4)}$ is a chiral projector.
A double variation $\delta_1^2$ on $y_{12}$ vanishes due to
two $\Sigma_+$ colliding. As we will be projecting to
the $\Theta=0$ component in the end
the resulting $\Theta_{12}$ from the variation
of $y_{12}$ must be compensated by the action of
$\delta_2$. The only relevant variation is thus
\[
e_{12}^\mu=\delta_1\delta_2 y_{12}^\mu=
-2\epsilon_1^\trans\tilde\Sigma_{+}P\Sigma^\mu \tilde\Sigma_-\epsilon_2.
\]
We can introduce an effective variation $\delta$ acting on $y_{12}$ by
\[
\delta y_{12}^\mu=e_{12}^\mu,\qquad
\delta \bar y_{12}^\mu=\bar e_{12}^\mu,
\]
with $\bar e_{12}^\mu=-2\epsilon_1^\trans\tilde\Sigma_{+}\bar P\Sigma^\mu \tilde\Sigma_-\epsilon_2$.
For $l$ consecutive variations we have to make sure the effective
variation produces the right combinatorial factors.
All variations on $z_1$ can be performed first, this yields $l$ powers 
of $\Theta_{12}$. The variations on $z_2$ should
later on annihilate all these and there are $l!$ equivalent ways to do so, 
thus we get
\[
\bigeval{(\delta_1\delta_2)^l F(z_{12})}_0=
l!\,\bigeval{\delta^l F(z_{12})}_0.
\]
For the variations of the scalar correlator we find
using \eqref{eq:scaprop}, \eqref{eq:proptop}
\<\label{eq:vary2pt}
%\bigeval{P_{12}^\Delta}_0\eq
%\frac{1}{|x_{12}|^{2\Delta}},
%\nln
\bigeval{(\delta_1\delta_2)^2 P_{12}^\Delta}_0\eq
\Delta(\Delta-2)\,
e^{ij}_{1}\,e^{kl}_2\,
\frac{\delta_{i[k}\delta_{l]j}}{|x_{12}|^{2\Delta+2}} +\ldots,
\nln
\bigeval{(\delta_1\delta_2)^4 P_{12}^\Delta}_0\eq
9\Delta^2(\Delta-2)^2(e^{i\mu}_{1} e^{j}_{1\mu})(e^{k\nu}_2 e^{l}_{2\nu})
\frac{\delta_{i(k}\delta_{l)j}}{|x_{12}|^{2\Delta+4}}+\ldots,
\>
for the components that correspond to 
BMN operators with two scalar impurities.
The complete set of even variations
can be found in App.~\ref{sec:propvary}.
The tensors $e_1^{ij}$, $e_2^{ij}$ are the
$\epsilon^2$ variation coefficients at the
two points $z_1$ and $z_2$, see \eqref{eq:epssqr}.
For the correlator of two BMN operators 
with $\Delta=\Delta^J_n-J$
we may write this in a general form as
\[\label{eq:BMN2ptgen}
\lreval{(\delta_1\delta_2)^l \bigvev{\OpB^{J}_n(z_1,V)\,\OpB^{J}_m(z_2,\bar V)}}_0
=
\sum_{a,b}
\bigbrk{N^{J,(l/2)}_{a,n}}^2 \bigbrk{\epsilon_1^l}^a \bigbrk{\epsilon_2^l}^b \frac{\delta_{nm}\,J_{12,ab}}{|x_{12}|^{2\Delta^J_n+l}},
\]
similar to \eqref{eq:BMNvargen}.
Here, the symbol $J_{12,ab}$ relates the spacetime and internal indices of the 
operators $a$ and $b$ in a conformally covariant way,
usually as a product of $J_{12,mn}=\delta_{mn}$ for 
$\grSO(6)$ indices and $J_{12,\mu\nu}=\eta_{\mu\nu}-2x_{12,\mu}x_{12,\nu}/x_{12}^2$ for
spacetime indices. 
We assume the descendant operators to be normalised such that
their two-point function is 
\[\label{eq:BMNdes2pt}
\lreval{\bigvev{\OpB^{J,(l/2)}_{a,n}(z_1,V)\,\OpB^{J,(l/2)}_{b,m}(z_2,\bar V)}}_0
=\frac{\delta_{nm}\,J_{12,ab}}{|x_{12}|^{2\Delta^J_n+l}}.
\]
Using the definition of descendant operators in \eqref{eq:BMNvargen}
we can compare 
\eqref{eq:BMN2ptgen}, \eqref{eq:BMNdes2pt}
to \eqref{eq:vary2pt} with $\Delta=\Delta^J_n-J\approx 2$ to read off 
the exact normalisation coefficients in terms of 
the scaling dimension. 
We obtain 
\<
N^{J,(1)}_{[ij],n}\eq\pm \sqrt{\Delta^J_n-J}\,\sqrt{\Delta^J_n-J-2}\approx 
\pm \sqrt{2}\times\sqrt{8}\,\sqrt{\frac{\gym^2 N}{8\pi^2}}\sin\frac{\pi n}{J+3},
\nln
N^{J,(2)}_{(ij),n}\eq\pm 3(\Delta^J_n-J)(\Delta^J_n-J-2)\approx 
\pm 6\times 8\,\frac{\gym^2 N}{8\pi^2}\sin^2\frac{\pi n}{J+3}
\>
to be compared to \eqref{eq:NormEx}.
We collect all even normalisation coefficients 
in App.~\ref{sec:norm}. 
They can easily be seen to agree 
with the direct variation of the bare
operators in App.~\ref{sec:BMNDesc} at leading order.
We note that some of the normalisation constants 
involve factors of $\delta\Delta^J_n=\Delta^J_n-J-2$,
the anomalous dimension of the BMN operators. 
In the direct variation of operators 
\eqref{eq:BMNvar} these correspond to 
factors of $(\kappa^J_n)^2\approx \delta\Delta^J_n$.

Moreover, this can be extended to higher genus.
Due to redefinitions of operators on the torus their variations 
can be altered by pieces proportional to $1/N^2$,
see \eqref{eq:varytorus}. 
These changes are reflected by a shift in
the scaling dimension on the torus. 
The anomalous dimension on the torus has been calculated in 
the BMN limit, see \eqref{eq:BMNdimtorus}.
Here, we obtain an exact normalisation constant 
$N^{J,(1)}_{[ij],n}=\sqrt{(\Delta^J_n-J)(\Delta^J_n-J-2)}
\sim \sqrt{2\delta\Delta^J_n}$
which agrees with \eqref{eq:varytorus}.
This, however, does not mean that supersymmetry 
determines the anomalous dimension on the torus. 
The crucial input of \eqref{eq:varytorus} is the mixing matrix 
which is the result of a $\order{\gtwo^2 \lambda'}$ calculation.
Supersymmetry only yields non-trivial relations among
the mixing matrix elements \cite{Beisert:2002bb} belonging 
to different descendant operators.

%%%%%%%%%%%%%%%%%%%%%%%%%%%%%%%%%%%%%%%%%%%%%%%%%%%%%%%%%%%%%%%%%%%%%%%%%%%%%
\subsection{Three-Point Functions}

In the following we will consider the implications
of superconformal symmetry on three-point functions 
of vacuum and BMN operators. 
A few correlators of descendants of these operators 
were computed in the BMN limit in 
\cite{Kristjansen:2002bb,Constable:2002hw,Chu:2002pd,Beisert:2002bb,Constable:2002vq}.
It can be observed that some of these correlators vanish and 
others are closely related to another. 
These relations will be explained.

The form of three-point functions of primary operators 
is in general not completely fixed by superconformal symmetry. 
We will therefore have to rely on some additional constraint to fix the form,
namely the $\half$ BPS condition of the vacuum operators $\OpV^J$. 
A general treatment of three-point function with some desired
properties is rather involved, see e.g. \cite{Osborn:1998qu}, and
we will have to make some simplifying assumptions here:
The operators under consideration are singlets under 
the spacetime group and carry internal vector indices. 
Consequently we will assume that a generic three-point function can 
be constructed from the building blocks $P$ and $K_{mn}$ that
have turned out useful before,
see also \cite{Eden:1999gh}. 
We will use three guiding principles in the construction.
The conformal dimensions at the three points must match the
dimensions of the operators, the indices must be saturated and
for vacuum operators the $\half$ BPS condition must hold 
manifestly by \eqref{eq:KBPS}.
With these guidelines we are able to explain the structure
of the three point functions that have been worked out explicitly.
The third principle, however, does not exclude the existence of
a three-point function that has the $\half$ BPS 
condition fulfilled by other means. 

%%%%%%%%%%%%%%%%%%%%%%%%%%%%%%%%%%%%%%
\paragraph{Three vacuum operators.}
To demonstrate the method we consider a three-point function of 
three vacuum operators \cite{Howe:1998zi,Eden:1999gh}.
This three-point function is unique due to three 
$\half$ BPS conditions to be satisfied. 
Effectively this means
that it depends only on $24$ instead of
$48$ fermionic coordinates. 
By superconformal transformations we can gauge away up to $32$ 
fermionic coordinates, which fixes this 
three-point function of scalar operators uniquely \cite{Eden:1999gh}.
Due to the $\half$ BPS conditions at all points the 
functions $K$ can only be used in the combinations
$K_{12}$, $K_{23}$, $K_{31}$.
Assume the charges of the operators are $J_1$, $J_2$, $J_3$. 
Then it is easily seen that
$(J_1+J_2-J_3)/2$ indices 
at point $z_1$ have to be connected to $z_2$ and so on. 
The only way to do this is in the combination $K_{12}$.
Multiplying the other two connections we
find that the dimensions of the operators match automatically.
The correlator is thus
\<
&&\bigvev{\OpV^{J_1}(z_1,V_1)\,\OpV^{J_2}(z_2,V_2)\,\OpV^{J_3}(z_3,V_3)}
=
\nl
\qquad\qquad C^{J_1J_2J_3} K_{12}^{(J_1+J_2-J_3)/2}
K_{23}^{(J_2+J_3-J_1)/2}
K_{31}^{(J_3+J_1-J_2)/2}.
\>
There is one condition that the charges must satisfy, namely
the powers of $K_{12},K_{23},K_{31}$ must be 
non-negative and integer \cite{Eden:1999gh}. 
Otherwise the 
$\grSO(6)$ indices cannot be fully saturated. 
Put differently, the $K$s must yield 
polynomial expressions in $V_1$, $V_2$. 
We refrain from giving an explicit coefficient $C^{J_1J_2J_3}$ for
this three-point function 
because the issue of 
diagonalisation of the vacuum sector is not settled
\cite{Beisert:2002bb}. 
For the bare operators and $J_1+J_2=J_3$
an all-genus expression is found in \cite{Kristjansen:2002bb}.

%%%%%%%%%%%%%%%%%%%%%%%%%%%%%%%%%%%%%%%%%%%%%%%%%%
\paragraph{Two vacuum and one BMN operator.}
Next we consider a three-point function of 
two vacuum operators at points $z_1,z_2$ and one BMN operator at point $z_3$.
Due to two $\half$ BPS conditions this three-point function
depends on $32$ fermionic coordinates and is 
apparently uniquely fixed as well, see above.
The $\half$ BPS conditions at points $z_1,z_2$ imply that the 
propagators can only be used in the combinations
$K_{12}$,
$V_1^m K_{13,mp}$ and 
$V_2^n K_{23,np}$.
The two last combinations can be contracted with $V_3^p$ as before
or they can be joined 
\[
K_{132}=V_1^m K_{13,m}{}^{p} K_{32,pn} V_1^n.
\]
This connection is 
different from the direct connection $K_{12}$
in that it has conformal dimension $2$ at point $z_3$.
In total there are four 
building blocks for this three-point function, 
$K_{12}$, $K_{13}$, $K_{23}$, $K_{132}$
and the resulting correlator is
\<
&&\bigvev{\OpV^{J_1}(z_1,V_1)\,\OpV^{J_2}(z_2,V_2)\,\OpB_n^{J_3}(z_3,V_3)}
=
\nl
\qquad\qquad
C^{J_1J_2J_3}_n
K_{132}^{a}
K_{12}^{b}
K_{13}^{(J_1+J_3-J_2)/2}
K_{23}^{(J_2+J_3-J_1)/2}.
\>
The numbers $a,b$ must be adjusted such that 
the conformal dimensions match and all 
$V$s come in the right power, i.e. 
$a=\half(\Delta^{J_3}_n-J_3)$, $b=\half(J_1+J_2-\Delta^{J_3}_n)$.
Three correlators of this form have been
calculated for descendant operators
in the BMN limit and at leading order 
\cite{Constable:2002hw,Chu:2002pd,Beisert:2002bb,Constable:2002vq} 
\<
&&\bigvev{\OpV^{J_1,[1]}_{i}(z_1,V)\,\OpV^{J_2,[1]}_{j}(z_2,V)\,
\OpB^{J_3}_{n}(z_3,\bar V)}=0,\quad\quad J_3=J_1+J_2-2,
\nl
\bigvev{\OpV^{J_1,[1]}_{i}(z_1,V)\,\OpV^{J_2,[1]}_{j}(z_2,V)\,
\OpB^{J_3,(1)}_{[kl],n}(z_3,\bar V)}=0,\quad J_3=J_1+J_2-3,
\nl
\bigvev{\OpV^{J_1,[1]}_{i}(z_1,V)\,\OpV^{J_2,[1]}_{j}(z_2,V)\,
\OpB^{J_3,(2)}_{(kl),n}(z_3,\bar V)}\neq 0,\quad J_3=J_1+J_2-4.
\>
The second zero is readily explained,  
$J_1+J_2+J_3$ is odd and therefore the
$\grSO(6)$ indices cannot be contracted. 
The first zero could be explained by
demanding that all $K$ come in non-negative
powers. Negative powers of $K$ would give rise
to non-polynomial expressions in the $V$s.
Strictly speaking, this is not a problem here, 
because $K_{12}$ and $K_{132}$ 
are both proportional to
$V_1\cdott V_2$ at $\Theta=0$.
Their product is a positive integer power of
$V_1\cdott V_2$ plus fermionic (nilpotent) 
corrections and thus still polynomial. 
Nevertheless we will use this positivity bound 
as a guiding principle in the construction 
of further three-point functions. 
It appears to give satisfactory results here, but it may 
be a wrong assumption in general. 
Positivity would imply
\[
J_1+J_2\geq \Delta^{J_3}_n.
\]
If this is not satisfied, the correlator should vanish.
%

%%%%%%%%%%%%%%%%%%%%%%%%%%%%%%%%%%%%%%%%%%%%%%%%%%
\paragraph{One vacuum and two BMN operators.}
The final and most interesting three-point function is a correlator of
two BMN operators at $z_1,z_2$ and one vacuum operator at $z_3$.
The single $\half$ BPS condition is not enough to fix the form of the function 
uniquely.
It allows the following six
building blocks $P_{12}$, $K_{12}$, $K_{13}$, $K_{23}$, 
$K_{123}$, $K_{213}$ and the three-point function is
\[\label{eq:BMN3pt}
\bigvev{\OpB_n^{J_1}(z_1,V_1)\,\OpB_m^{J_2}(z_2,V_2)\,\OpV^{J_3}(z_3,V_3)}
=\sum_a C^{J_1J_2J_3}_{nm}(a)\,
P_{12}^a
K_{12}^b
K_{13}^c
K_{123}^d
K_{23}^e
K_{213}^f.
\]
The exponents are related by 
the charge conservation relations
$b+c+d=J_1$, 
$b+e+f=J_2$,
$c+d+e+f=J_3$ and
conformal dimension matching relations
$a+b+c+d+2f=\Delta^{J_1}_n$,
$a+b+e+2d+f=\Delta^{J_2}_m$.
This yields 
\<
b\eq \half(J_1+J_2-J_3)
\nln
c\eq\half(J_1+J_3-\Delta^{J_2}_m+a)
\nln
d\eq\half(\Delta^{J_2}_m-J_2-a)
\nln
e\eq\half(J_2+J_3-\Delta^{J_1}_n+a)
\nln
f\eq\half(\Delta^{J_1}_n-J_1-a)
\>
with one free parameter $a$. For a general combination of $J_1, J_2, J_3$ 
the form of the three-point function is not fixed by superconformal symmetry.
Taking derivatives with respect to $a$ should yield (nilpotent)
invariants,
each of which gives rise to an independent structure constant $C$.
The case that we are going to consider is
an extremal correlator with $J_1=J_2+J_3$
where we might hope for another constraint.
Assuming that $e$ and $f$ must be positive,
the form of this extremal correlator would be uniquely fixed,
because $e+f=0$ with the only solution $e=f=0$ and 
\[
a=\Delta^{J_1}_n-J_1\qquad \mbox{for }J_1=J_2+J_3.
\]
The single structure constant in the BMN limit could then be read off
from the correlator of the singlet operator in \cite{Beisert:2002bb}
\[\label{eq:BMN3ptConst}
C^{J,Jr,J(1-r)}_{nm}\sim\frac{2\gtwo\sqrt{1-r}\,\sin^2(\pi n r)}
{\sqrt{Jr}\,\pi^2\bigbrk{n^2-\sfrac{m^2}{r^2}}^2}
\times\frac{m^2}{r^2}.
\]

We now set $V_2=V_3=\bar V_1$ and perform the variations 
$(\delta_1\delta_2)^l$ on \eqref{eq:BMN3pt}. 
On the left hand side we get the descendant operators
multiplied by their 
normalisation constants 
$N^{J,(l/2)}_{a,n}$ and
$N^{Jr,(l/2)}_{b,m}$.
On the right hand side we first act
with $\delta_1^l$. This affects only $P_{12}$.
It gives rise to $l$ powers of 
$\Theta_{12}$ which need to be saturated
by $\delta_2^l$ when we set $\Theta=0$. 
Therefore $(\delta_1\delta_2)^l$ effectively acts 
only on $P_{12}$.
This results in the same expressions as for the two-point 
function \eqref{eq:BMN2ptgen} of the operator $\OpB^{J}_{n}$ and 
a factor of $\bigbrk{N^{J,(l/2)}_{a,n}}^2$.
The generic descendant correlator
for normalised operators is thus
\<\label{eq:vierachtunddreissig}
&&
\bigeval{\bigvev{\OpB^{J,(l/2)}_{a,n}(z_1,V)\,
\OpB^{Jr,(l/2)}_{b,m}(z_2,\bar V)\,\OpV^{J(1-r)}(z_3,\bar V)}}_0
\\\nonumber&&
\qquad\qquad\qquad=
\frac{\bigbrk{N^{J,(l/2)}_{a,n}/N^{Jr,(l/2)}_{b,m}}\, C^{J,Jr,J(1-r)}_{mn}\, J_{12,ab}}
{|x_{12}|^{\Delta^J_n+\Delta^{Jr}_m-J(1-r)+l}\,
|x_{23}|^{\Delta^{Jr}_m+J(1-r)-\Delta^{J}_n}\,|x_{13}|^{\Delta^{J}_n+J(1-r)-\Delta^{Jr}_m}}.
\>
The normalisation constants $N^{J,(l)}_{a,n}$ can be found in App.~\ref{sec:norm}.
In the BMN limit and to leading order 
the quotient of normalisation constants can only be 
$1$, $nr/m$ or $n^2r^2/m^2$
which effectively replaces 
the last factor in \eqref{eq:BMN3ptConst} by 
$m^2/r^2$, $mn/r$ and $n^2$, respectively.
One should keep in mind that this form of the function is based 
on some assumptions. 
In principle, one should analyse all three-point covariants 
relevant to these operators, a task beyond the scope of this work. 
It needs to be compared to 
some explicit calculations. For instance it does
coincide with the correlators 
of antisymmetric and symmetric-traceless BMN
operators in \cite{Beisert:2002bb,Constable:2002vq}.

%%%%%%%%%%%%%%%%%%%%%%%%%%%%%%%%%%%%%%%%%%%%%%%%%%
\paragraph{Three BMN operators.}
A three-point function of three BMN operators is less 
constrained than the 
above three-point functions and none of these correlators has
been calculated so far. We will therefore not investigate it here.

%%%%%%%%%%%%%%%%%%%%%%%%%%%%%%%%%%%%%%%%%%%%%%%%%%%%%%%%%%%%%%%%%%%%%%%%%%%%%
\section{Discussion and Outlook}\label{sec:Concl}

In this paper we have investigated the implications of 
superconformal symmetry on the BMN operators with
two charge defects. 
It was seen how the BMN operators fill out
multiplets of $\grSU(4)$ and $\grSU(2\mathord{,}2|4)$
and an abstract way of defining them was found.
We have determined the form of the operators 
with scalar impurities at finite charge $J$
and their anomalous dimension.
This was then used to derive the form of all
other bosonic operators by supersymmetry.
Finally, we have presented superspace three-point functions 
involving BMN operators which agree with previously 
computed correlators.
\medskip

The main result of the group theoretical considerations
is that the long supermultiplets of $\superN=4$ super Yang-Mills
theory whose lowest dimensional operators are 
scalars of naive dimension $J+2$ transforming in the irreducible representation
$[0,J,0]$ of $\grSU(4)$ contain exactly the BMN operators with two impurities. 
This is an alternative definition of these operators as opposed to
the heuristic construction of BMN. The new definition
enables one to apply useful results 
of superconformal symmetry and representation theory
of $\grSU(2\mathord{,}2|4)$ to the set of BMN operators. 

We have found the exact one-loop, 
planar form of the BMN operators at finite $J$,
see \eqref{eq:BMNScalar}.
The obtained phase factors differ 
from any of the previously conjectured ones. 
This is without consequences in the BMN limit
because the modification is negligible. 
It is, however, crucial when $J$ is small. 
We find that some of the operators with small $J$ 
coincide with operators which 
have been intensely studied in recent years. 
Most importantly this is the Konishi operator at $J=0$, $n=1$.
Furthermore, some of the dimension four 
operators studied in \cite{Bianchi:2002rw} coincide
with our expressions.
We are able to reproduce their anomalous dimension
with a single expression that remains valid in the BMN limit.
Furthermore, it can be seen that the mixing pattern of single-trace and 
double-trace operators \cite{Beisert:2002bb} persists 
at small $J$, see for example \cite{Bianchi:1999ge,Arutyunov:2000ku,Bianchi:2002rw}.
We thus found a class of operators that interpolates between 
the BMN limit and operators at low dimension.
\medskip

In fact, it seems to be more useful to classify 
operators of $\superN=4$ SYM by their number of 
defects than by their dimension. 
The larger the number of 
defects, the further the operators are `away' from the
protected operators. As the number of impurities increases
the number of operators vastly increases and 
operator mixing becomes more and more complicated.
In contrast, increasing the charge of an operator while
keeping the number of defects constant adds only a 
manageable amount of combinatorics to the problem.
The additional charged fields act as an inert background 
to the original operator in many respects.

In this proposed classification a single-trace 
operator is characterised by several numbers.
The most important ones are the number of 
defects, $k$, and the $\grSO(4)$ and $\grSO(3,1)$ representations. 
The total spin of these representations is bounded 
from above by $k-2$. 
Furthermore, there is the $\grSO(2)$ charge $J$.
If the charge is reasonably large 
compared to the number of defects, 
the defects can be viewed as a dilute gas
and one should expect that everything
depends `smoothly' on $J$. 
As proposed in \cite{Berenstein:2002jq}
the operators with single-charge defects 
($\phi_i,\cder_\mu,\psi$)
will then organise themselves
in terms of $k-1$ mode numbers.
In such a way the spectrum of
strings on a plane-wave is obtained.
The operators with multiple-charge defects 
($\bar Z,F_{\mu\nu},\bar \psi,\ldots$),
which seemingly do not fit into the string spectrum,
were expected \cite{Berenstein:2002jq} to become infinitely massive 
in the BMN limit and decouple from the low-lying modes. 
Interestingly, we find that in the case of two defects 
this does not happen. 
Nevertheless, if for a different reason, the agreement of spectra is not 
spoiled by the presence of additional insertions.
Operators with multiple-charge defects are hidden within the 
ordinary ones with single-charge defects by operator mixing. 
They give rise to an additional mode 
instead of an exceptional operator.
Therefore all operators are classified by a single mode number
and their correlation functions depend `smoothly' on it. 
One might conjecture that this holds in general. 
We have also seen that the same mode decomposition of
operators can be extended all the way down to
the smallest possible charge $J$. 
(Certainly, only low winding numbers would be allowed for these
and the term `mode decomposition' becomes somewhat inappropriate).
Again, this might hold in general.
Then the remaining characteristic numbers 
are given by a set of $k-1$ mode numbers. 
In that case, the spectrum would be very similar to the spectrum 
of a single string in free string theory. 
Including multiple-trace operators
we would naturally arrive at an interacting string theory. 
This exhausts the spectrum of local operators in the gauge theory.
Consequently, this characterisation might be very appealing
for the general AdS/CFT correspondence away from the plane-wave limit. 

In this context the BMN limit of an arbitrary operator could 
be obtained by taking the charge of the operator to infinity
while keeping all other classifying numbers fixed. 
In the case of operators with two defects we have seen that 
the mode decomposition includes all operators. 
There are no exceptional operators which become infinitely massive
as suggested in~\cite{Berenstein:2002jq}.
It follows that the complete set of operators 
(with a finite mode number) survives in the BMN limit.
Although we have not found any infinitely massive operators
in the BMN limit so far, it might well be that they exist 
among the operators with more than two defects.
If so, the classification scheme would have to be enhanced accordingly.%
\medskip

We note that the form of the primary BMN operator \eqref{eq:PrimaryOp}
is only the lowest order approximation to its
full form. Redefinitions are required
at higher genus and higher loops.
We have illustrated the modifications that
occur at genus one, higher corrections
will be similar.
Beyond one-loop one should expect mixing
between operators with scalar, 
fermionic and derivative insertions.
The diagonalisation would involve a redefinition of the 
primary operators $\OpB^J_n$ by 
operators with equal quantum numbers
($\OpB^{J-2,(2)}_k$, $\OpB^{J-4,(4)}_k$).
One may argue that this is negligible in 
the BMN limit: Due to the dilute gas property
interactions involving both impurities should be suppressed. 
This is true in the planar limit, but not in general.
The additional pieces in the amplitudes of
singlet operators in \cite{Beisert:2002bb} 
are exclusively due to interactions between 
both impurities.
We should stress that in this work we have mostly been considering 
full, diagonalised operators. 
If the correspondence to strings
on plane-waves is true, one should find that 
their anomalous dimensions agree with the 
eigenvalues of the string Hamiltonian. 
Current attempts to compare both theories 
\cite{Parnachev:2002kk,Gross:2002mh,Vaman:2002ka,Pearson:2002zs,Gomis:2002wi}
do not try to accomplish that, however. 
They aim at comparing matrix elements 
at the level of bare operators/states. 
\medskip

In order to show that all the BMN operators belong to the same 
supermultiplet we have worked out 
supersymmetry descendants of the primary operator,
see \eqref{eq:BMNvar}.
We have also shown how supersymmetry relates
the mixing matrices. 
This reduces the complexity of future calculations, 
as only the primary operator has to be taken into account.
Supersymmetry is then used to derive the corresponding statements
for the descendants.
The form of the descendants in App.~\ref{sec:BMNDesc} can be used as a dictionary.

Using superconformal symmetry we have found two-point and three-point functions
of BMN operators.
We have worked out the two-point correlators
for descendants operators, see \eqref{eq:vary2pt} and obtained 
the exact expressions for the normalisation coefficients 
of the variation of the operators. 
A complete set of correlators and normalisation constants is found in
App.~\ref{sec:propvary} and~\ref{sec:norm}. 
Three-point functions of BMN operators 
were presented, most importantly
\eqref{eq:vierachtunddreissig}. They explain the relations between 
correlators that have been determined recently
\cite{Beisert:2002bb,Constable:2002vq}.
In principle, these should enable one to derive
expressions for a large class of descendant correlators. 
\bigskip

Several questions concerning 
BMN operators at finite charge and their 
classification suggest themselves.
An extension of the current analysis to
operators with more than two defects
would be interesting.
In particular an enumeration of
such operators and their 
explicit form at one-loop and at the planar level
might lead to new insights into the proposed classification.
For instance, one might expect some
new features to appear at the level of four defects. 
The representation $[0,J,0]$ with dimension $J+4$
is not on the unitary bound and its constituents 
might therefore behave quite differently.
For example their anomalous dimensions are not required to be positive. 
This representation might also lead to the simplest examples of operators
that become infinitely massive in the BMN limit, should these
exist at all.
Furthermore, in the low charge regime we would expect to find 
the operators with $J=0$ investigated in \cite{Arutyunov:2002rs}.
Alternatively, one could work in the opposite direction and 
try to generalise some results involving low-dimensional operators,
like the Konishi operators,
to arbitrary charge and to the BMN limit.
A two-loop generalisation of some of the results involving 
BMN operators at finite $J$ would also be useful.
For example, the anomalous dimension depends on
two parameters, $J$ and $n$. 
By investigating the dependence of 
the two-loop result on the additional parameters 
one might be able to guess the structure 
of the higher-loop anomalous dimensions.
In that sense the BMN operators and the BMN limit might lead
to a better understanding of 
$\superN=4$ SYM and the AdS/CFT correspondence in general.

%%%%%%%%%%%%%%%%%%%%%%%%%%%%%%%%%%%%%%%%%%%%%%%%%%%%%%%%%%%%%%%%%%%%%%%%%%%%%
\subsection*{Acknowledgements}

The author would like to thank 
Gleb Arutyunov,
Thomas Klose,
Stefano Kovacs,
Charlotte Kristjansen,
Ari Pankiewicz and
Jan Plefka 
for useful discussions.
In particular I am grateful to 
Matthias Staudacher 
for 
many helpful suggestions and comments 
on the manuscript.

\newpage
%%%%%%%%%%%%%%%%%%%%%%%%%%%%%%%%%%%%%%%%%%%%%%%%%%%%%%%%%%%%%%%%%%%%%%%%%%%%%
\appendix
%%%%%%%%%%%%%%%%%%%%%%%%%%%%%%%%%%%%%%%%%%%%%%%%%%%%%%%%%%%%%%%%%%%%%%%%%%%%%
\section{Spinors in $D=9+1$}
\label{sec:spinors}   
 
We use indices the $M,N,\ldots=0,\ldots,9$ for vectors
and indices $A,B,\ldots=1,\ldots,16$ for spinors.
Some $\grSO(9,1)$ invariant tensors
are the metric $\eta_{MN}=\diag(-,+,\ldots,+)$,
the antisymmetric tensor $\eps_{MNOPQRSTUV}$,
and the tensors
$\Sigma^M_{AB}$ and
$\tilde\Sigma_M^{AB}$
relating two spinor indices with one vector index.
The $\Sigma$ matrices (where the
spinor indices are commonly suppressed) 
satisfy the Clifford algebra
\[
\Sigma_M\tilde\Sigma_N+\Sigma_N\tilde\Sigma_M=2\eta_{MN}.
\]

\noindent\textbf{Normal ordering}
\<\label{eq:sigmaalg}
\Sigma_M\tilde\Sigma_N\eq\Sigma_{MN}+\eta_{MN},
\nln
\Sigma_M\tilde\Sigma_N\Sigma_R\eq
\Sigma_{MNR}+\eta_{MN}\Sigma_R
 -\eta_{MR}\Sigma_N+\eta_{NR}\Sigma_M,
\nln
\Sigma_M\tilde\Sigma_N\Sigma_R\tilde\Sigma_S\eq
\Sigma_{MNRS}
\nl
+\eta_{MN}\Sigma_{RS}
-\eta_{MR}\Sigma_{NS}
+\eta_{MS}\Sigma_{NR}
\nl
+\eta_{NR}\Sigma_{MS}
-\eta_{NS}\Sigma_{MR}
+\eta_{RS}\Sigma_{MN}
\nl
+\eta_{MN}\eta_{RS}
-\eta_{MR}\eta_{NS}
+\eta_{MS}\eta_{NR}
%\nln
%\ldots
\>
where $\Sigma_{MNR\ldots}$ is defined to be the 
antisymmetrised (`normal ordered') product of $\Sigma$s 
\[
\Sigma_{MNR\ldots}=\Sigma_{[M}\tilde\Sigma_{N}\Sigma_{R}\cdots
\]
and tilded symbols are obtained by all
$\Sigma$s replaced by $\tilde\Sigma$s and vice versa.
\medskip

\noindent\textbf{Chisholm identities} (for a Clifford algebra of $d$ 
$\Sigma_i$ matrices)
\<
\Sigma_i\Sigma_{(n)}\Sigma^i
\eq (-1)^n (d-2n) \Sigma_{(n)},
\nln
\Sigma_{ij}\Sigma_{(n)}\Sigma^{ij}
\eq \bigbrk{d-(d-2n)^2} \Sigma_{(n)},
\nln
\Sigma_{ijk}\Sigma_{(n)}\Sigma^{ijk}
\eq (-1)^n(d-2n)\bigbrk{3d-2-(d-2n)^2} \Sigma_{(n)},
\\\nonumber&&\ldots
\>
We have suppressed the tilde on every
second $\Sigma$ and $\Sigma_{(n)}$ denotes 
an normal ordered product of $n$ $\Sigma_i$s.
This rule is also applicable when a subset of 
$d$ of the $D=9+1$ $\Sigma$s is considered.
\medskip

\noindent
\textbf{Symmetries}
\<
\Sigma_M^\trans\eq+\Sigma_M=\Sigma_M^\conj=\Sigma_M^\dagger,
\nln
\Sigma_{MN}^\trans\eq -\tilde\Sigma_{MN},
\nln
\Sigma_{MNP}^\trans\eq -\Sigma_{MNP},
\nln
\Sigma_{MNPQ}^\trans\eq +\tilde\Sigma_{MNPQ},
\\\nonumber&&\ldots
\>
\textbf{Dualisations} (opposite signs for $\tilde \Sigma$s)
\<
1\eq +\sfrac{1}{10!}\eps_{MNOPQRSTUV}\Sigma^{MNOPQRSTUV}
\nln
\Sigma_{M}\eq -\sfrac{1}{9!}\eps_{MNOPQRSTUV}\Sigma^{NOPQRSTUV}
\nln
\Sigma_{MN}\eq -\sfrac{1}{8!}\eps_{MNOPQRSTUV}\Sigma^{OPQRSTUV}
\nln
\Sigma_{MNO}\eq +\sfrac{1}{7!}\eps_{MNOPQRSTUV}\Sigma^{PQRSTUV}
\\\nonumber&&\ldots
\>
\textbf{Fierz identities}
\<
a^{[A} b^{B]}
\eq \sfrac{1}{16\cdott 3!}(a^\trans\Sigma^{MNP} b)\tilde\Sigma^{AB}_{MNP} 
\nln
a_{[A} b_{B]}
\eq \sfrac{1}{16\cdott 3!}(a^\trans\tilde\Sigma^{MNP} b)\Sigma_{AB}^{MNP} 
\nln
a^{\{A} b^{B\}}
\eq \sfrac{1}{16}(a^\trans\Sigma^{M} b)\tilde\Sigma^{AB}_{M} 
+\sfrac{1}{16\cdott 2\cdott 5!}(a^\trans\Sigma_{MNOPQ} b)\tilde\Sigma^{AB}_{MNOPQ} 
\nln
a_{\{A} b_{B\}}
\eq \sfrac{1}{16}(a^\trans\tilde\Sigma_{M} b)\Sigma_{AB}^{M} 
+\sfrac{1}{16\cdott 2\cdott 5!}(a^\trans\tilde\Sigma_{MNOPQ} b)\Sigma_{AB}^{MNOPQ} 
\nln
a^{A} b_{B}
\eq 
\sfrac{1}{16}(a^\trans b)\delta^A_B
+\sfrac{1}{16\cdott 2!}(a^\trans \Sigma^{MN} b)\tilde\Sigma_{MN}{}^A{}_B 
+\sfrac{1}{16\cdott 4!}(a^\trans \Sigma^{MNPQ} b)\tilde\Sigma_{MNPQ}{}^A{}_B
\>

\paragraph{Reduction to $D=3+1$, $\superN=4$.}
To be able to work with this 
notation in $\superN=4$ SYM we need to split 
up the vectors in $4$ and $6$ components.
The spacetime vectors are labelled by indices
$\mu,\nu,\ldots=0,\ldots,3$ and
the internal vectors by 
$m,n,\ldots=1,\ldots 6$.
Due to this $\grSO(3,1)\times \grSO(6)$ split we obtain two 
antisymmetric invariant tensors
$\eps_{\mu\nu\rho\sigma}$ and $\eps_{mnpqrs}$
and one can build an invariant combination of $\Sigma$s
\<
\Sigma_{(4)}=\sfrac{1}{4!}\eps_{\mu\nu\rho\sigma}\Sigma^{\mu\nu\rho\sigma}
\eq -\sfrac{1}{6!}\eps_{mnopqr}\Sigma^{mnopqr}=-\Sigma_{(6)}
\nln
\tilde\Sigma_{(4)}=\sfrac{1}{4!}\eps_{\mu\nu\rho\sigma}\tilde\Sigma^{\mu\nu\rho\sigma}
\eq +\sfrac{1}{6!}\eps_{mnopqr}\tilde\Sigma^{mnopqr}=+\tilde\Sigma_{(6)}
\>
with the properties
\[
\Sigma_{(4)}^2=\Sigma_{(6)}^2=\tilde\Sigma_{(4)}^2=\tilde\Sigma_{(6)}^2=-1,\quad
\Sigma_{(4)}^\trans=\tilde\Sigma_{(4)},\quad
\Sigma_{(6)}^\trans=-\tilde\Sigma_{(6)}.
\]
These give rise to chiral projectors
\[
P=\half+\sfrac{i}{2}\Sigma_{(4)},\quad
\bar P=\half-\sfrac{i}{2}\Sigma_{(4)}.
\]
Useful identities for the computation of traces involving the projectors are
\[
P\Sigma_{\mu\nu\rho\sigma}=i\eps_{\mu\nu\rho\sigma}P,\qquad
P\Sigma_{mnopqr}=i\eps_{mnopqr}P.
\]

\paragraph{Harmonic coordinates.}
We introduce a complex internal vector $V_m$ with the 
properties $V^2=0$, $|V|^2=1$. 
Two components of an internal vector are
specialised by this, $a_+=a\cdot V$, $a_-=a\cdot \bar V$.
The remaining four are labelled by indices 
$i,j,\ldots=1,\ldots,4$.
There are two projectors
\[
P_+=\half \Sigma_-\tilde\Sigma_+,\qquad
P_-=\half \Sigma_+\tilde\Sigma_-
\]
which effectively project to 
spinors of the $\grSO(7,1)$ subgroup 
of $\grSO(9,1)$ that leaves $V$ and $\bar V$ invariant. 

\newpage
%%%%%%%%%%%%%%%%%%%%%%%%%%%%%%%%%%%%%%%%%%%%%%%%%%%%%%%%%%%%%%%%%%%%%
\section{Descendant operators}\label{sec:BMNDesc}

In this appendix we present all bosonic BMN operators with two defects
and how they are related by supersymmetry transformations
that do not change the number of defects.\medskip

\noindent\textbf{Definitions}
\<
\delta\eq
\epsilon^\trans\tilde\Sigma_+ D
\nln
e_{MN}\eq
\epsilon^\trans\tilde\Sigma_{+MN}\epsilon
\nln
\kappa_n^J\eq
\sqrt{8}\,\sqrt{\frac{\gym^2 N}{8\pi^2}}\,\sin\frac{\pi n}{J+3}\sim
\sqrt{\lambda'}\,n
\nln
e_{M+}\eq e_{M-}=0
\>

\noindent\textbf{Fierz identities} 

\noindent level 2
\[
\tilde\Sigma_+\epsilon\epsilon^\trans\tilde\Sigma_+=
\sfrac{1}{16}e_{MN}
\tilde\Sigma_+\Sigma^{MN}
\]

\noindent level 4
\<
e^{ij}e^{kl}\eq 
e^{[ij}e^{kl]}
+e^{[k}{}_{\mu}\delta^{l][j}e^{i]\mu}
+\sfrac{1}{6}\delta^{i[k}\delta^{l]j}e^{mn}e_{mn},
\nln
e^{ij}e^{k\mu}\eq
e^{[ij}e^{k]\mu}
+\sfrac{2}{3}\delta^{k[i}e^{j]\nu}e^{\mu}{}_{\nu},
\nln
e^{i\mu}e^{j \nu}\eq
-e^{ij}e^{\mu\nu}
+\sfrac{1}{4}\eta^{\mu\nu}e^{i\rho}e^{j}{}_{\rho}
+\sfrac{1}{4}\delta^{ij}e^{\mu k}e^{\nu}{}_{k},
\nln
e^{i k}e^{j}{}_{k}\eq
\half e^{i\mu}e^{j}{}_{\mu}+\quarter\delta^{ij}e^{lk}e_{lk},
\nln
e^{ik}e^{\mu}{}_{k}\eq e^{i\rho}e^{\mu}{}_{\rho},
\nln
e^{i\mu}e_{i\mu}\eq 0,
\nln
e^{ij}e_{ij}\eq -e^{\mu\nu}e_{\mu\nu},
\nln
e_{[MN}e_{OP]}\eq -\sfrac{1}{4!}\eps_{+-MNOPQRST}e^{[QR}e^{ST]},
\>

\noindent level 6
\<
e^{i\mu}e_{k\mu}e^{kj}\eq e^{ij}e^{kl}e_{kl}
\nln
e^{ik}e_{lk}e^{lj}\eq \sfrac{3}{4}e^{ij}e^{kl}e_{kl}
\>

\noindent level 8
\[e^{i\mu}e_{j\mu}e^{j\nu}e_{i\nu}= 2e^{ij}e_{ij}e^{kl}e_{kl}\]
\pagebreak[5]

\noindent\textbf{Variations of \mathversion{bold}$\OpB^{J}_n$}
\<\label{eq:BMNvarall}
\delta^2 \OpB^{J}_n\eq 
\sqrt{2}\,e^{ij}\, \kappa_n^J\,\OpB^{J,(1)}_{ij,n}
+2\sqrt{2}\,e^{\mu\nu}\,\OpB^{J,(1)}_{\mu\nu,n}
+2\sqrt{2}\,e^{\mu i}\,\OpB^{J,(1)}_{\mu i,n}
\nln
\delta^4\OpB^J_n\eq
-6\bigbrk{\kappa^J_n}^2\,e^{i\mu}e^{j}{}_{\mu}\,\OpB^{J,(2)}_{(ij),n}
+24 e^{\mu i}e^{\nu}{}_i\,\OpB^{J,(2)}_{(\mu\nu),n}
+16\sqrt{2}\,\kappa^J_n\, e^{\mu j}e^{i}{}_j\,\OpB^{J,(2)}_{\mu i,n}
\nl
+4\sqrt{2}\,\kappa^J_n \,e^{ij}e_{ij}\,\OpB^{J,(2)}_{n}
+4\sqrt{6}\,\kappa^J_n \,e^{[MN}e^{RS]}\,\OpB^{J,(2)}_{[MNRS]}
\nl
-2e^{jk}e_{jk}
\frac{N_0^{-J-2}}{\sqrt{J+3}}\cos\frac{\pi n}{J+3}
\Tr \bigbrk{\cder^\mu\cder_\mu Z+[i\phi^i,[i\phi_i,Z]]+i\psi^\trans \Sigma_+\psi} Z^{J+1}
\nln
\delta^6\OpB^{J}_{n}\eq
120\sqrt{2}\,\bigbrk{\kappa^J_n}^2\, e^{ij}e^{kl}e_{kl}\,\OpB^{J,(3)}_{[ij],n} 
+240\sqrt{2}\,\kappa^J_n\, e^{\mu\nu}e^{\rho\sigma}e^{\rho\sigma}\,\OpB^{J,(3)}_{[\mu \nu],n}
\nl
-160\sqrt{2}\,\bigbrk{\kappa^J_n}^2\, e^{\mu j}e_{\nu j}e^{\nu i}\,\OpB^{J,(3)}_{\mu i,n}+\mbox{EOM}
\nln
\delta^8\OpB^{J}_{n}\eq -2240\bigbrk{\kappa^J_n}^2\, e^{ij}e_{ij}e^{kl}e_{kl}\,\OpB^{J,(4)}_n
+\mbox{EOM}
\>

\noindent\textbf{Variations of \mathversion{bold}$\OpB^{J,(2)}_n$}
\<
\delta^2\OpB^{J,(2)}_{n}\eq
e^{ij}\,\kappa^J_n\, \OpB^{J,(3)}_{[ij],n}
-2e^{\mu\nu}\,\OpB^{J,(3)}_{[\mu\nu],n},
\nln
\delta^4\OpB^{J,(2)}_{n}\eq
-4\sqrt{2}\,\kappa^J_n\,e^{ij}e_{ij}\,\OpB^{J,(4)}_{n}
\>

%\delta^2\OpB^{J,(2)}_{i\mu,n}\eq
%-2e_{ij}\,\kappa^J_n\,\OpB^{J,(3)}_{\mu j}
%+2e_{\mu \nu}\,\kappa^J_n\, \OpB^{J,(3)}_{\nu i}
%+4e_{i\nu}\,\OpB^{J,(3)}_{[\mu\nu],n}
%+2e^{\mu j}\,\kappa^J_n\, \OpB^{J,(3)}_{[ij],n}
%\nln
%\delta^2\OpB^{J,(2)}_{(ij),n}\eq
%-2\sqrt{2}e^{i\mu}\OpB^{J,(3)}_{\mu j,n}
%-2\sqrt{2}e^{j\mu}\OpB^{J,(3)}_{\mu i,n}
%+2\sqrt{2}e^{ik}\OpB^{J,(3)}_{[kj],n}
%+2\sqrt{2}e^{jk}\OpB^{J,(3)}_{[ki],n}
%\nln
%\delta^2\OpB^{J,(2)}_{(\mu\nu),n}\eq
%+\sqrt{2} e_{\nu \rho}\kappa^J_n \OpB^{J,(3)}_{\mu\rho,n}
%+\sqrt{2} e_{\mu \rho}\kappa^J_n \OpB^{J,(3)}_{\nu\rho,n}
%-\sfrac{1}{\sqrt{2}}e_{\nu i}(\kappa^J_n)^2\OpB^{J,3}_{\mu i,n}
%-\sfrac{1}{\sqrt{2}}e_{\mu i}(\kappa^J_n)^2\OpB^{J,3}_{\nu i,n}
%\nln
%\delta^2\OpB^{J,(2)}_{[MNRS],n}\eq
%-\sfrac{6}{\sqrt{3}}\kappa^J_n e_{[MN}\delta_{R}^i\delta_{S]}^j
%\OpB^{J,(3)}_{[ij],n}
%+\sfrac{12}{\sqrt{3}} \kappa^J_n  e_{[MN}\delta_{R}^\mu\delta_{S]}^i
%\OpB^{J,(3)}_{\mu i,n}
%-\sfrac{12}{\sqrt{3}}e_{[MN}\delta_{R}^\mu\delta_{S]}^\nu
%\OpB^{J,(3)}_{[\mu\nu]}
%\nln
%\delta^2
%\OpB^{J,(3)}_{\mu i}\eq
%2\sqrt{2}\,e_{\mu i}\,\OpB^{J,(4)}_{n}
%\nln
%\delta^2\OpB^{J,(3)}_{[\mu\nu],n}\eq
%-\sqrt{2}\,e_{\mu\nu}\,\kappa^J_n\,\OpB^{J,(4)}_n
%\nln
%\delta^2\OpB^{J,(3)}_{[ij],n}\eq
%-2\sqrt{2}\,e_{ij}\,\OpB^{J,(4)}_{n}
%\nln

\noindent\textbf{Descendant Operators}

\noindent{level 0}
\<
\OpB^J_n\eq
\frac{N_0^{-J-2}}{\sqrt{J+3}}
\bigg[
\half\sum_{p=0}^J\cos\frac{\pi n(2p+3)}{J+3}
\Tr \phi_i Z^p \phi_i Z^{J-p}
\nl
\qquad\qquad\qquad
-2\cos\frac{\pi n}{J+3}\Tr \bar Z Z^{J+1}
\bigg]
\>

\noindent{level 2}
\<
\OpB^{J,(1)}_{[ij],n}\eq
\frac{N_0^{-J-3}}{\sqrt{J+3}}
\sum_{p=0}^{J+1}
i\sin\frac{\pi n(2p+2)}{J+3}\Tr \phi_{[i}Z^p\phi_{j]} Z^{J+1-p}
\nln
\OpB^{J,(1)}_{[\mu\nu],n}\eq
\frac{N_0^{-J-2}}{\sqrt{J+3}}\bigg[\sfrac{1}{8\sqrt{2}}\sum_{p=0}^J \cos \frac{\pi n(2p+3)}{J+3}
\Tr \psi^\trans  Z^p \Sigma_{+\mu\nu}\psi Z^{J-p}
\nl
\qquad\qquad\qquad+\sfrac{1}{\sqrt{2}}\cos\frac{\pi n}{J+3}\Tr F_{\mu\nu} Z^{J+1}
\bigg]
\nln
\OpB^{J,(1)}_{\mu i,n}\eq
\frac{N_0^{-J-2}}{\sqrt{J+3}}\bigg[\sfrac{1}{\sqrt{2}}\sum_{p=0}^J \cos \frac{\pi n(2p+3)}{J+3}
\Tr  \phi_i Z^p   \cder_\mu Z Z^{J-p}
\nl
\qquad\qquad\qquad+\sqrt{2}\,\cos\frac{\pi n}{J+3}\Tr  \cder_\mu\phi_i Z^{J+1}
\nl
\qquad\qquad+\sfrac{1}{8\sqrt{2}}\sum_{p=0}^J \cos \frac{\pi n(2p+3)}{J+3}
\Tr \psi^\trans  Z^p \Sigma_{+\mu i}\psi Z^{J-p}
\bigg]
\>

\noindent{level 4}
\<
\OpB^{J,(2)}_{(ij),n}\eq
\frac{N_0^{-J-4}}{\sqrt{J+3}}
\sum_{p=0}^{J+2}
\cos\frac{\pi n(2p+1)}{J+3}\Tr \phi_{(i}Z^p\phi_{j)} Z^{J+2-p}
\nln
\OpB^{J,(2)}_{(\mu\nu),n}\eq
\frac{N_0^{-J-2}}{\sqrt{J+3}}\bigg[\half\sum_{p=0}^J \cos\frac{\pi n (2p+3)}{J+3}\Tr \cder_{(\mu} Z Z^p \cder_{\nu)} Z Z^{J-p}
\nl
\qquad\qquad\qquad+\half\cos\frac{\pi n}{J+3}\Tr \cder_{(\mu} \cder_{\nu)} Z Z^{J+1}
\bigg]
\nln
\OpB^{J,(2)}_{\mu i,n}\eq
\frac{N_0^{-J-3}}{\sqrt{J+3}}\sum_{p=0}^{J+1} i\sin\frac{\pi n(2p+2)}{J+3}
\Tr \phi_i Z^{p} \cder_\mu Z Z^{J+1-p}
\nln
\OpB^{J,(2)}_{n}\eq
\frac{N_0^{-J-3}}{\sqrt{J+3}}\bigg[\sfrac{1}{8}\sum_{p=0}^{J+1}i\sin\frac{\pi n(2p+2)}{J+3}
\Tr \psi^\trans Z^p \Sigma_+ \psi Z^{J+1-p}
\nl
\qquad\qquad+\sfrac{1}{2}
\sum_{p=0}^{J+1}i\sin\frac{\pi n(2p+2)}{J+3}
\Tr \phi_i Z^p [i\phi_i,Z] Z^{J+1-p}
\bigg]
\nln
%\OpB^{J,(2)}_{n}\eq
%\frac{N_0^{-J-3}}{\sqrt{J+3}}\bigg[\sfrac{1}{8}\sum_{p=0}^{J+1}i\sin\frac{\pi n(2p+2)}{J+3}
%\Tr \psi^\trans Z^p \Sigma_+ \psi Z^{J+1-p}
%\nl
%\qquad\qquad-\sin\frac{\pi n}{J+3}
%\sum_{p=0}^{J+2}\cos\frac{\pi n(2p+1)}{J+3}
%\Tr \phi_i Z^p \phi_i Z^{J+2-p}
%\bigg]
%\nln
\OpB^{J,(2)}_{[MNRS],n}\eq
\frac{N_0^{-J-3}}{32\sqrt{3}\,\sqrt{J+3}}
\sum_{p=0}^{J+1}i\sin\frac{\pi n(2p+2)}{J+3}
\Tr \psi^\trans Z^p \Sigma_{+MNRS}\psi Z^{J+1-p}
\>

\noindent{level 6}
\<
\OpB^{J,(3)}_{[\mu\nu],n}\eq
\frac{N_0^{-J-3}}{\sqrt{J+3}}\bigg[\half\sum_{p=0}^{J+1}
i\sin\frac{\pi n(2p+2)}{J+3}
\Tr \cder_{[\mu} Z  Z^p \cder_{\nu]} Z Z^{J+1-p}
\nl
\qquad\qquad-\sfrac{1}{16}
\sum_{p=0}^{J+1}
i\sin\frac{\pi n(2p+2)}{J+3}
\Tr \psi^\trans \Sigma_{+\mu\nu}  Z^p [i\psi,Z] Z^{J+1-p}
\bigg]
\nln
%\OpB^{J,(3)}_{[\mu\nu],n}\eq
%\frac{N_0^{-J-3}}{\sqrt{J+3}}\bigg[\half\sum_{p=0}^{J+1}
%i\sin\frac{\pi n(2p+2)}{J+3}
%\Tr \cder_\mu Z  Z^p \cder_\nu Z Z^{J+1-p}
%\nl
%\qquad\qquad+\sfrac{1}{8}\sin\frac{\pi n}{J+3}
%\sum_{p=0}^{J+2}
%\cos\frac{\pi n(2p+1)}{J+3}
%\Tr \psi^\trans \Sigma_{+\mu\nu}  Z^p\psi Z^{J+2-p}
%\bigg]
%\nln
\OpB^{J,(3)}_{[ij],n}\eq
\frac{N_0^{-J-4}}{\sqrt{J+3}}
\bigg[
\sfrac{1}{8\sqrt{2}}\sum_{p=0}^{J+2}
\cos\frac{\pi n(2p+1)}{J+3}
\Tr \psi^\trans \Sigma_{+ij}  Z^p\psi Z^{J+2-p}
\nl
\qquad\qquad+\sfrac{1}{\sqrt{2}}
\sum_{p=0}^{J+2}
\cos\frac{\pi n(2p+1)}{J+3}
\Tr \phi_i Z^{p} [i\phi_j,Z] Z^{J+2-p}
\bigg]
\nln
\OpB^{J,(3)}_{\mu i}\eq
\frac{N_0^{-J-4}}{\sqrt{J+3}}
\bigg[
\sfrac{1}{\sqrt{2}}\sum_{p=0}^{J+2} \cos\frac{\pi n(2p+1)}{J+3}\Tr \phi_i Z^{p} \cder_\mu Z Z^{J+2-p}
\nl
\qquad\qquad-
\sfrac{1}{8\sqrt{2}}\sum_{p=0}^{J+2}\cos\frac{\pi n(2p+1)}{J+3}
\Tr \psi^\trans \Sigma_{+\mu i}  Z^{p} \psi Z^{J+2-p}
\bigg]
\>

\noindent{level 8}
\<
\OpB^{J,(4)}_{n}\eq
\frac{N_0^{-J-4}}{\sqrt{J+3}}\bigg[
\sfrac{1}{4}\sum_{p=0}^{J+2}\cos\frac{\pi n(2p+1)}{J+3}
\Tr \cder^\mu Z Z^{p} \cder_\mu Z Z^{J+2-p}
\nl
\qquad\qquad
-\sfrac{1}{8}\sum_{p=0}^{J+2}\cos\frac{\pi n(2p+1)}{J+3}
\Tr \psi^\trans\Sigma_+ Z^{p} [i\psi,Z] Z^{J+2-p}
\nl
\qquad\qquad
+\sfrac{1}{4}\sum_{p=0}^{J+2}\cos\frac{\pi n(2p+1)}{J+3}
\Tr [i\phi_i,Z]   Z^{p} [i\phi_i,Z] Z^{J+2-p}
\bigg]
\>

%%%%%%%%%%%%%%%%%%%%%%%%%%%%%%%%%%%%%%%%%%%%%%%%%%%%%%%%%%%%
\section{Variations of the scalar superspace propagator}
\label{sec:propvary}

In this appendix we present 
all components of the scalar 
superspace propagator corresponding to bosonic BMN 
operators with two defects.\medskip

\noindent\textbf{Definitions}
\<
P_{12}\eq \frac{1}{|y_{12}|^\Delta|\bar y_{12}|^\Delta}
\nln
J_{12,\mu\nu} \eq
\eta_{\mu\nu}-2\,\frac{x_{12,\mu}x_{12,\nu}}{x_{12}^2},\qquad
\nln
J_{12,mn}\eq \delta_{mn}.
\nln
y_{12}^\mu
\eq x_{12}^\mu+\sfrac{i}{2}\Theta_{12}^\trans \Sigma_{(4)}\Sigma^\mu\Theta_{12},
\nln
e_{12}^\mu\eq 
-2\epsilon_1^\trans\tilde\Sigma_{+}P\Sigma^\mu \tilde\Sigma_-\epsilon_2.
\>

\noindent\textbf{Variations of half a scalar propagator}
\<
\delta\frac{1}{y_{12}^{\Delta}}\eq
-\Delta\frac{y_{12}\cdott e_{12}}{|y_{12}|^{\Delta+2}}
\nln
\delta^2\frac{1}{y_{12}^{\Delta}}\eq
\Delta\frac{(\Delta+2)(y_{12}\cdott e_{12})^2-y_{12}^2 e_{12}^2}{|y_{12}|^{\Delta+4}}
\nln
\delta^3\frac{1}{y_{12}^{\Delta}}\eq
-\Delta(\Delta+2)\frac{(y_{12}\cdott e_{12})\bigbrk{(\Delta+4)(y_{12}\cdott e_{12})^2-3y_{12}^2e_{12}^2}}{|y_{12}|^{\Delta+6}}
\\\nonumber
\delta^4\frac{1}{y_{12}^{\Delta}}\eq
\Delta(\Delta+2)\frac{(\Delta+6)(
\Delta+4)(y_{12}\cdott e_{12})^4
-6(\Delta+4)y_{12}^2e_{12}^2(y_{12}\cdott e_{12})^2
+3y_{12}^4e_{12}^4
}{|y_{12}|^{\Delta+8}}
\>

\noindent\textbf{Fierz identities}
\<
\bigeval{(y_{12}\cdott e_{12})^2}_0\eq
\quarter x_{12}^2\delta_{ik}\delta_{jl}\, e^{ij}_{1}e^{kl}_2
-\sfrac{i}{8} x_{12}^2\eps_{+-ijkl}\,e^{ij}_1 e^{kl}_2
\nl
+\quarter x_{12}^2 J_{12,\mu\rho}J_{12,\nu\sigma}\, e^{\mu\nu}_1 e^{\rho\sigma}_2
-\sfrac{i}{8}x_{12}^2 J_{12,\mu\kappa}J_{12,\nu\lambda}\eps^{\kappa\lambda}{}_{\rho\sigma}\,
 e^{\mu\nu}_1 e^{\rho\sigma}_2
\nln
\bigeval{e_{12}^2}_0\eq
\delta_{ik}\delta_{jl}\,e^{ij}_1 e^{kl}_2-\sfrac{i}{2}\eps_{+-ijkl}\,e^{ij}_{1}e^{kl}_{2}
\nln
\bigeval{(e_{12}\cdott y_{12}) (\bar e_{12}\cdott \bar y_{12})}_0\eq
\half x_{12}^2 J_{12,\mu\nu}\delta_{ij}\, e^{\mu i}_1 e^{\nu j}_2
\>

\noindent\textbf{Variations of the propagator}\nopagebreak

\noindent{level 0}
\[
\bigeval{P_{12}^\Delta}_0=
\frac{1}{|x_{12}|^{2\Delta}}
\]

\noindent{level 2}
\<
\bigeval{(\delta_1\delta_2)^2 P_{12}^\Delta}_0\eq
\Delta(\Delta-2)
e^{ij}_{1}e^{kl}_2
\frac{\delta_{i[k}\delta_{l]j}}{|x_{12}|^{2\Delta+2}} 
\nl
+\Delta(\Delta+2)
e^{\mu\nu}_{1}e^{\rho\sigma}_{2}
\frac{J_{12,\mu[\rho}J_{21,\sigma]\nu}}{|x_{12}|^{2\Delta+2}} 
\nl
+2\Delta^2 
e^{\mu i}_{1}e^{\nu j}_2
\frac{J_{12,\mu\nu}\delta_{ij}}{|x_{12}|^{2\Delta+2}}
\>

\noindent{level 4}
\<
\bigeval{(\delta_1\delta_2)^4 P_{12}^\Delta}_0\eq
9\Delta^2(\Delta-2)^2(e^{i\mu}_{1} e^{j}_{1\mu})(e^{k\nu}_2 e^{l}_{2\nu})
\frac{\delta_{i(k}\delta_{l)l}}{|x_{12}|^{2\Delta+4}}
\nl
+9\Delta^2(\Delta+2)^2
(e^{\mu i}_1 e^{\nu}_{1i}) (e^{\rho j}_2  e^{\sigma}_{2j})
\frac{J_{12,\mu(\rho}J_{21,\sigma)\nu}}{|x_{12}|^{2\Delta+4}}
\nl
+32\Delta^2(\Delta+2)(\Delta-2)(e^{ik}_{1}e^{\mu}_{1k})(e^{\nu l}_2 e^{j}_{2l})
\frac{J_{12,\mu\nu}\delta_{ij}}{|x_{12}|^{2\Delta+4}},
\nl
+2\Delta^2(\Delta+2)(\Delta-2)(e^{ij}_1e_{1ij})(e^{kl}_2e_{2kl})
\frac{1}{|x_{12}|^{2\Delta+4}},
\nlnum
+6\Delta^2(\Delta+2)(\Delta-2)(e^{[MN}_1e^{OP]}_1)(e^{[QR}_2e^{ST]}_2)
\frac{J_{12,MQ}J_{12,NR}J_{12,OS}J_{12,PT}}{|x_{12}|^{2\Delta+4}}
\>

\noindent{level 6}
\<
\bigeval{(\delta_1\delta_2)^6 P_{12}^\Delta}_0\eq
900\Delta^3(\Delta+2)(\Delta-2)^2
(e^{ij}_{1}e^{mn}_1e_{1mn})(e^{ij}_2e^{rs}_2e_{2rs})
\frac{\delta_{i[k}\delta_{l]j}}{|x_{12}|^{2\Delta+6}}
\nl
+900\Delta^3(\Delta+2)^2(\Delta-2)
(e^{\mu\nu}_{1}e^{\kappa\lambda}_1e_{1\kappa\lambda})(e^{\rho\sigma}_2e^{\tau\pi}_2e_{2\tau\pi})
\frac{J_{12,\mu[\rho}J_{21,\sigma]\nu}}{|x_{12}|^{2\Delta+6}}
\nl
+800\Delta^2(\Delta+2)^2(\Delta-2)^2
(e^{\mu k}_1e_{1\rho k}e^{\rho i}_1)(e^{\nu l}_2e_{2\sigma l}e^{\sigma j}_2)
\frac{J_{12,\mu\nu}\delta_{ij}}{|x_{12}|^{2\Delta+6}}
\>

\noindent{level 8}
\[
\bigeval{(\delta_1\delta_2)^8 P_{12}^\Delta}_0=
19600\Delta^4(\Delta+2)^2(\Delta-2)^2
(e^{ij}_1e_{1ij}e^{kl}_1e_{1kl})(e^{mn}_1e_{1mn}e^{rs}_1e_{1rs})
\frac{1}{|x_{12}|^{2\Delta+8}}
\]
%

%\newpage
%%%%%%%%%%%%%%%%%%%%%%%%%%%%%%%%%%%%%%%%%%%%%%%%%%%%%%%%%%%%%%%%%%%%%
\section{Normalisation coefficients}\label{sec:norm}

We present the normalisation coefficients
of the supersymmetry variations of BMN
operators. The variations can be written 
in a general form as
\[
\delta^l \OpB^J_n=\sum_a N^{J,(l/2)}_{a,n} (\epsilon^l)^a \OpB^{J,(l/2)}_{a,n}.
\]
The variation parameters $(\epsilon^l)^a$ are those from
the previous appendices and the operators are canonically 
normalised.
The the right hand side of the following table 
can be read off from App.~\ref{sec:propvary}.
The left hand side is from App.~\ref{sec:BMNDesc},
it is the one-loop planar approximation of
the full result.
The anomalous dimension is $\delta\Delta^J_n=\Delta^J_n-J-2$.
\<
1=\mathord{}&N^{J}_n&\mathord{}=1,
\nln
\sqrt{2}\,\kappa^J_n\approx\mathord{}&N^{J,(1)}_{[ij],n}&\mathord{}=
(\delta\Delta^J_n)^{1/2}(2+\delta\Delta^J_n)^{1/2},
\nln
2\sqrt{2}\approx\mathord{}&N^{J,(1)}_{[\mu\nu],n}&\mathord{}=
(2+\delta\Delta^J_n)^{1/2}(4+\delta\Delta^J_n)^{1/2},
\nln
2\sqrt{2}\approx\mathord{}&N^{J,(1)}_{\mu i,n}&\mathord{}=
\sqrt{2}\,(2+\delta\Delta^J_n),
\nln
-6(\kappa^J_n)^2\approx\mathord{}&N^{J,(2)}_{(ij),n}&\mathord{}=
-3\delta\Delta^J_n(2+\delta\Delta^J_n),
\nln
24\approx\mathord{}&N^{J,(2)}_{(\mu\nu),n}&\mathord{}=
3(2+\delta\Delta^J_n)(4+\delta\Delta^J_n),
\nln
16\sqrt{2}\,\kappa^J_n\approx\mathord{}&N^{J,(2)}_{\mu i,n}&\mathord{}=
4\sqrt{2}\,(\delta\Delta^J_n)^{1/2}(2+\delta\Delta^J_n)(4+\delta\Delta^J_n)^{1/2},
\nln
4\sqrt{2}\,\kappa^J_n\approx\mathord{}&N^{J,(2)}_n&\mathord{}=
\sqrt{2}\,(\delta\Delta^J_n)^{1/2}(2+\delta\Delta^J_n)(4+\delta\Delta^J_n)^{1/2},
\nln
4\sqrt{6}\,\kappa^J_n\approx\mathord{}&N^{J,(2)}_{[MNRS],n}&\mathord{}=
\sqrt{6}(\delta\Delta^J_n)^{1/2}(2+\delta\Delta^J_n)(4+\delta\Delta^J_n)^{1/2},
\nln
120\sqrt{2}(\kappa^J_n)^2\approx\mathord{}&N^{J,(3)}_{[ij],n}&\mathord{}=
 30(\delta\Delta^J_n)(2+\delta\Delta^J_n)^{3/2}(4+\delta\Delta^J_n)^{1/2}, 
\nln
240\sqrt{2}\kappa^J_n\approx\mathord{}&N^{J,(3)}_{[\mu\nu],n}&\mathord{}=
30(\delta\Delta^J_n)^{1/2}(2+\delta\Delta^J_n)^{3/2}(4+\delta\Delta^J_n),
\nln
-160\sqrt{2}(\kappa^J_n)^2\approx\mathord{}&N^{J,(3)}_{\mu i,n}&\mathord{}=
- 20\sqrt{2}\,(\delta\Delta^J_n)(2+\delta\Delta^J_n)(4+\delta\Delta^J_n),
\nln
-2240(\kappa^J_n)^2\approx\mathord{}&N^{J,(4)}_n&\mathord{}= 
-140(\delta\Delta^J_n)(2+\delta\Delta^J_n)^2(4+\delta\Delta^J_n).
\>

%%%%%%%%%%%%%%%%%%%%%%%%%%%%%%%%%%%%%%%%%%%%%%%%%%%%%%%%%%%%%%%%%

%\bibliography{BMNSConf}

\begin{thebibliography}{10}
\raggedright
\small
\parskip 0pt
\itemsep 4pt

%%CITATION = HEP-TH 0202021;%%
\bibitem{Berenstein:2002jq}
D.~Berenstein, J.~M.~Maldacena and H.~Nastase,
\textit{``Strings in flat space and pp waves from {$\mathcal{N}=\mathord{}$4}
  {Super} {Yang Mills}''},
JHEP~\textbf{04}~(2002)~013,
\hypref{http://arXiv.org/abs/hep-th/0202021}{\texttt{hep-th/0202021}}.

%%CITATION = HEP-TH 0201081;%%
\bibitem{Blau:2002dy}
M.~Blau, J.~Figueroa-O'Farrill, C.~Hull and G.~Papadopoulos,
\textit{``Penrose limits and maximal supersymmetry''},
Class.~Quant.~Grav.~\textbf{19}~(2002)~L87,
\hypref{http://arXiv.org/abs/hep-th/0201081}{\texttt{hep-th/0201081}}.

%%CITATION = HEP-TH 0112044;%%
\bibitem{Metsaev:2001bj}
R.~R.~Metsaev,
\textit{``Type {IIB} {Green-Schwarz} superstring in plane wave {Ramond-Ramond}
  background''},
Nucl.~Phys.~\textbf{B625}~(2002)~70,
\hypref{http://arXiv.org/abs/hep-th/0112044}{\texttt{hep-th/0112044}}.

%%CITATION = HEP-TH 0204146;%%
\bibitem{Spradlin:2002ar}
M.~Spradlin and A.~Volovich,
\textit{``Superstring interactions in a pp-wave background''},
Phys.~Rev.~\textbf{D66}~(2002)~086004,
\hypref{http://arXiv.org/abs/hep-th/0204146}{\texttt{hep-th/0204146}}.

%%CITATION = HEP-TH 0205174;%%
\bibitem{Gopakumar:2002dq}
R.~Gopakumar,
\textit{``String interactions in PP-waves''},
Phys.~Rev.~Lett.~\textbf{89}~(2002)~171601,
\hypref{http://arXiv.org/abs/hep-th/0205174}{\texttt{hep-th/0205174}}.

%%CITATION = HEP-TH 0206065;%%
\bibitem{Lee:2002rm}
P.~Lee, S.~Moriyama and J.~Park,
\textit{``Cubic interactions in pp-wave light cone string field theory''},
Phys.~Rev.~\textbf{D66}~(2002)~085021,
\hypref{http://arXiv.org/abs/hep-th/0206065}{\texttt{hep-th/0206065}}.

%%CITATION = HEP-TH 0206073;%%
\bibitem{Spradlin:2002rv}
M.~Spradlin and A.~Volovich,
\textit{``Superstring interactions in a pp-wave background. {II}''},
JHEP~\textbf{01}~(2003)~036,
\hypref{http://arXiv.org/abs/hep-th/0206073}{\texttt{hep-th/0206073}}.

%%CITATION = HEP-TH 0208179;%%
\bibitem{Schwarz:2002bc}
J.~H.~Schwarz,
\textit{``Comments on superstring interactions in a plane-wave background''},
JHEP~\textbf{09}~(2002)~058,
\hypref{http://arXiv.org/abs/hep-th/0208179}{\texttt{hep-th/0208179}}.

%%CITATION = HEP-TH 0208209;%%
\bibitem{Pankiewicz:2002gs}
A.~Pankiewicz,
\textit{``More comments on superstring interactions in the pp-wave
  background''},
JHEP~\textbf{09}~(2002)~056,
\hypref{http://arXiv.org/abs/hep-th/0208209}{\texttt{hep-th/0208209}}.

%%CITATION = HEP-TH 0210246;%%
\bibitem{Pankiewicz:2002tg}
A.~Pankiewicz and B.~Stefanski,~Jr.,
\textit{``PP-Wave Light-Cone Superstring Field Theory''},
\hypref{http://arXiv.org/abs/hep-th/0210246}{\texttt{hep-th/0210246}}.

%%CITATION = HEP-TH 0205033;%%
\bibitem{Kristjansen:2002bb}
C.~Kristjansen, J.~Plefka, G.~W.~Semenoff and M.~Staudacher,
\textit{``A new double-scaling limit of {$\mathcal{N}=\mathord{}$4} super
  {Yang-Mills} theory and {PP}-wave strings''},
Nucl.~Phys.~\textbf{B643}~(2002)~3,
\hypref{http://arXiv.org/abs/hep-th/0205033}{\texttt{hep-th/0205033}}.

%%CITATION = HEP-TH 0205066;%%
\bibitem{Gross:2002su}
D.~J.~Gross, A.~Mikhailov and R.~Roiban,
\textit{``Operators with large R charge in {$\mathcal{N}=\mathord{}$4}
  Yang-Mills theory''},
Annals~Phys.~\textbf{301}~(2002)~31,
\hypref{http://arXiv.org/abs/hep-th/0205066}{\texttt{hep-th/0205066}}.

%%CITATION = HEP-TH 0205089;%%
\bibitem{Constable:2002hw}
N.~R.~Constable, D.~Z.~Freedman, M.~Headrick, S.~Minwalla, L.~Motl,
  A.~Postnikov and W.~Skiba,
\textit{``{PP}-wave string interactions from perturbative {Yang-Mills}
  theory''},
JHEP~\textbf{07}~(2002)~017,
\hypref{http://arXiv.org/abs/hep-th/0205089}{\texttt{hep-th/0205089}}.

%%CITATION = HEP-TH 0205270;%%
\bibitem{Arutyunov:2002xd}
G.~Arutyunov and E.~Sokatchev,
\textit{``Conformal fields in the pp-wave limit''},
JHEP~\textbf{08}~(2002)~014,
\hypref{http://arXiv.org/abs/hep-th/0205270}{\texttt{hep-th/0205270}}.

%%CITATION = HEP-TH 0205321;%%
\bibitem{Bianchi:2002rw}
M.~Bianchi, B.~Eden, G.~Rossi and Y.~S.~Stanev,
\textit{``On operator mixing in {$\mathcal{N}=\mathord{}$4} {SYM}''},
Nucl.~Phys.~\textbf{B646}~(2002)~69,
\hypref{http://arXiv.org/abs/hep-th/0205321}{\texttt{hep-th/0205321}}.

%%CITATION = HEP-TH 0206005;%%
\bibitem{Chu:2002pd}
C.-S.~Chu, V.~V.~Khoze and G.~Travaglini,
\textit{``Three-point functions in {$\mathcal{N}=\mathord{}$4} {Yang-Mills}
  theory and pp-waves''},
JHEP~\textbf{06}~(2002)~011,
\hypref{http://arXiv.org/abs/hep-th/0206005}{\texttt{hep-th/0206005}}.

%%CITATION = HEP-TH 0206079;%%
\bibitem{Santambrogio:2002sb}
A.~Santambrogio and D.~Zanon,
\textit{``Exact anomalous dimensions of {$\mathcal{N}=\mathord{}$4} Yang-Mills
  operators with large R charge''},
Phys.~Lett.~\textbf{B545}~(2002)~425,
\hypref{http://arXiv.org/abs/hep-th/0206079}{\texttt{hep-th/0206079}}.

%%CITATION = HEP-TH 0206248;%%
\bibitem{Huang:2002yt}
M.-X.~Huang,
\textit{``{String} interactions in pp-wave from {$\mathcal{N}=\mathord{}$4}
  {Super Yang Mills}''},
Phys.~Rev.~\textbf{D66}~(2002)~105002,
\hypref{http://arXiv.org/abs/hep-th/0206248}{\texttt{hep-th/0206248}}.

%%CITATION = HEP-TH 0208010;%%
\bibitem{Parnachev:2002kk}
A.~Parnachev and A.~V.~Ryzhov,
\textit{``Strings in the near plane wave background and {AdS/CFT}''},
JHEP~\textbf{10}~(2002)~066,
\hypref{http://arXiv.org/abs/hep-th/0208010}{\texttt{hep-th/0208010}}.

%%CITATION = HEP-TH 0208041;%%
\bibitem{Gursoy:2002yy}
U.~G{\"u}rsoy,
\textit{``Vector operators in the {BMN} correspondence''},
\hypref{http://arXiv.org/abs/hep-th/0208041}{\texttt{hep-th/0208041}}.

%%CITATION = HEP-TH 0208178;%%
\bibitem{Beisert:2002bb}
N.~Beisert, C.~Kristjansen, J.~Plefka, G.~W.~Semenoff and M.~Staudacher,
\textit{``BMN correlators and operator mixing in {$\mathcal{N}=\mathord{}$4}
  super Yang-Mills theory''},
Nucl.~Phys.~\textbf{B650}~(2003)~125,
\hypref{http://arXiv.org/abs/hep-th/0208178}{\texttt{hep-th/0208178}}.

%%CITATION = HEP-TH 0209002;%%
\bibitem{Constable:2002vq}
N.~R.~Constable, D.~Z.~Freedman, M.~Headrick and S.~Minwalla,
\textit{``Operator mixing and the BMN correspondence''},
JHEP~\textbf{10}~(2002)~068,
\hypref{http://arXiv.org/abs/hep-th/0209002}{\texttt{hep-th/0209002}}.

%%CITATION = HEP-TH 0209244;%%
\bibitem{Eynard:2002df}
B.~Eynard and C.~Kristjansen,
\textit{``BMN correlators by loop equations''},
JHEP~\textbf{10}~(2002)~027,
\hypref{http://arXiv.org/abs/hep-th/0209244}{\texttt{hep-th/0209244}}.

%%CITATION = HEP-TH 0209263;%%
\bibitem{Janik:2002bd}
R.~A.~Janik,
\textit{``BMN operators and string field theory''},
Phys.~Lett.~\textbf{B549}~(2002)~237,
\hypref{http://arXiv.org/abs/hep-th/0209263}{\texttt{hep-th/0209263}}.

%%CITATION = HEP-TH 0301150;%%
\bibitem{Klose:2003tw}
T.~Klose,
\textit{``Conformal Dimensions of Two-Derivative BMN Operators''},
JHEP~\textbf{03}~(2003)~012,
\hypref{http://arXiv.org/abs/hep-th/0301150}{\texttt{hep-th/0301150}}.

%%CITATION = HEP-TH 0206059;%%
\bibitem{Verlinde:2002ig}
H.~Verlinde,
\textit{``Bits, matrices and {1/N}''},
\hypref{http://arXiv.org/abs/hep-th/0206059}{\texttt{hep-th/0206059}}.

%%CITATION = HEP-TH 0208231;%%
\bibitem{Gross:2002mh}
D.~J.~Gross, A.~Mikhailov and R.~Roiban,
\textit{``A calculation of the plane wave string Hamiltonian from
  {$\mathcal{N}=\mathord{}$4} super-Yang-Mills theory''},
\hypref{http://arXiv.org/abs/hep-th/0208231}{\texttt{hep-th/0208231}}.

%%CITATION = HEP-TH 0209215;%%
\bibitem{Vaman:2002ka}
D.~Vaman and H.~Verlinde,
\textit{``Bit strings from {$\mathcal{N}=\mathord{}$4} gauge theory''},
\hypref{http://arXiv.org/abs/hep-th/0209215}{\texttt{hep-th/0209215}}.

%%CITATION = HEP-TH 0210102;%%
\bibitem{Pearson:2002zs}
J.~Pearson, M.~Spradlin, D.~Vaman, H.~Verlinde and A.~Volovich,
\textit{``Tracing the String: BMN correspondence at Finite $J^2/N$''},
\hypref{http://arXiv.org/abs/hep-th/0210102}{\texttt{hep-th/0210102}}.

%%CITATION = HEP-TH 0210153;%%
\bibitem{Gomis:2002wi}
J.~Gomis, S.~Moriyama and J.~Park,
\textit{``SYM description of SFT Hamiltonian in a pp-wave background''},
\hypref{http://arXiv.org/abs/hep-th/0210153}{\texttt{hep-th/0210153}}.

%%CITATION = HEP-TH 9712074;%%
\bibitem{Minwalla:1998ka}
S.~Minwalla,
\textit{``Restrictions imposed by superconformal invariance on quantum field
  theories''},
Adv. Theor.~Math.~Phys.~\textbf{2}~(1998)~781,
\hypref{http://arXiv.org/abs/hep-th/9712074}{\texttt{hep-th/9712074}}.

%%CITATION = HEP-TH 0201145;%%
\bibitem{Arutyunov:2002ff}
G.~Arutyunov and E.~Sokatchev,
\textit{``Implications of superconformal symmetry for interacting (2,0) tensor
  multiplets''},
Nucl.~Phys.~\textbf{B635}~(2002)~3,
\hypref{http://arXiv.org/abs/hep-th/0201145}{\texttt{hep-th/0201145}}.

%%CITATION = LMPHD,9,287;%%
\bibitem{Dobrev:1985vh}
V.~K.~Dobrev and V.~B.~Petkova,
\textit{``On the group theoretical approach to extended conformal
  supersymmetry: classification of multiplets''},
Lett.~Math.~Phys.~\textbf{9}~(1985)~287.

%%CITATION = HEP-TH 9812067;%%
\bibitem{Andrianopoli:1998ut}
L.~Andrianopoli and S.~Ferrara,
\textit{``Short and long SU(2,2/4) multiplets in the AdS/CFT correspondence''},
Lett.~Math.~Phys.~\textbf{48}~(1999)~145,
\hypref{http://arXiv.org/abs/hep-th/9812067}{\texttt{hep-th/9812067}}.

%%CITATION = HEP-TH 0104016;%%
\bibitem{Bianchi:2001cm}
M.~Bianchi, S.~Kovacs, G.~Rossi and Y.~S.~Stanev,
\textit{``Properties of the Konishi multiplet in {$\mathcal{N}=\mathord{}$4}
  SYM theory''},
JHEP~\textbf{05}~(2001)~042,
\hypref{http://arXiv.org/abs/hep-th/0104016}{\texttt{hep-th/0104016}}.

%%CITATION = NUPHA,B291,172;%%
\bibitem{Gates:1987is}
S.~J.~Gates,~Jr. and S.~Vashakidze,
\textit{``On D$\mathord{}=\mathord{}$10, {$\mathcal{N}=\mathord{}$1}
  supersymmetry, superspace geometry and superstring effects''},
Nucl.~Phys.~\textbf{B291}~(1987)~172.

%%CITATION = NUPHA,B136,461;%%
\bibitem{Sohnius:1978wk}
M.~F.~Sohnius,
\textit{``Bianchi identities for supersymmetric gauge theories''},
Nucl.~Phys.~\textbf{B136}~(1978)~461.

%%CITATION = NUPHA,B163,519;%%
\bibitem{Gates:1980jv}
S.~J.~Gates,~Jr. and W.~Siegel,
\textit{``Understanding constraints in superspace formulations of
  supergravity''},
Nucl.~Phys.~\textbf{B163}~(1980)~519.

%%CITATION = NUPHA,B169,347;%%
\bibitem{Gates:1980wg}
S.~J.~Gates,~Jr., K.~S.~Stelle and P.~C.~West,
\textit{``Algebraic origins of superspace constraints in supergravity''},
Nucl.~Phys.~\textbf{B169}~(1980)~347.

%%CITATION = HEP-TH 9906188;%%
\bibitem{Bianchi:1999ge}
M.~Bianchi, S.~Kovacs, G.~Rossi and Y.~S.~Stanev,
\textit{``On the logarithmic behavior in {$\mathcal{N}=\mathord{}$4} {SYM}
  theory''},
JHEP~\textbf{08}~(1999)~020,
\hypref{http://arXiv.org/abs/hep-th/9906188}{\texttt{hep-th/9906188}}.

%%CITATION = HEP-TH 0005182;%%
\bibitem{Arutyunov:2000ku}
G.~Arutyunov, S.~Frolov and A.~C.~Petkou,
\textit{``Operator product expansion of the lowest weight CPOs in
  {$\mathcal{N}=\mathord{}$4} SYM(4) at strong coupling''},
Nucl.~Phys.~\textbf{B586}~(2000)~547,
\hypref{http://arXiv.org/abs/hep-th/0005182}{\texttt{hep-th/0005182}}.

\bibitem{Todorov:1978rf}
I.~T.~Todorov, M.~C.~Mintchev and V.~B.~Petkova,
\textit{``Conformal invariance in quantum field theory''},
Sc. Norm. Sup. (1978),
Pisa, Italy,
273p.

%%CITATION = HEP-TH 9605009;%%
\bibitem{Erdmenger:1997yc}
J.~Erdmenger and H.~Osborn,
\textit{``Conserved currents and the energy-momentum tensor in conformally
  invariant theories for general dimensions''},
Nucl.~Phys.~\textbf{B483}~(1997)~431,
\hypref{http://arXiv.org/abs/hep-th/9605009}{\texttt{hep-th/9605009}}.

%%CITATION = HEP-TH 9808041;%%
\bibitem{Osborn:1998qu}
H.~Osborn,
\textit{``{$\mathcal{N}=\mathord{}$1} superconformal symmetry in
  four-dimensional quantum field theory''},
Annals~Phys.~\textbf{272}~(1999)~243,
\hypref{http://arXiv.org/abs/hep-th/9808041}{\texttt{hep-th/9808041}}.

%%CITATION = HEP-TH 9903230;%%
\bibitem{Park:1999pd}
J.-H.~Park,
\textit{``Superconformal symmetry and correlation functions''},
Nucl.~Phys.~\textbf{B559}~(1999)~455,
\hypref{http://arXiv.org/abs/hep-th/9903230}{\texttt{hep-th/9903230}}.

%%CITATION = HEP-TH 9808162;%%
\bibitem{Howe:1998zi}
P.~S.~Howe, E.~Sokatchev and P.~C.~West,
\textit{``3-point functions in {$\mathcal{N}=\mathord{}$4} Yang-Mills''},
Phys.~Lett.~\textbf{B444}~(1998)~341,
\hypref{http://arXiv.org/abs/hep-th/9808162}{\texttt{hep-th/9808162}}.

%%CITATION = HEP-TH 9905085;%%
\bibitem{Eden:1999gh}
B.~Eden, P.~S.~Howe and P.~C.~West,
\textit{``Nilpotent invariants in {$\mathcal{N}=\mathord{}$4} SYM''},
Phys.~Lett.~\textbf{B463}~(1999)~19,
\hypref{http://arXiv.org/abs/hep-th/9905085}{\texttt{hep-th/9905085}}.

%%CITATION = HEP-TH 0206020;%%
\bibitem{Arutyunov:2002rs}
G.~Arutyunov, S.~Penati, A.~C.~Petkou, A.~Santambrogio and E.~Sokatchev,
\textit{``Non-protected operators in {$\mathcal{N}=\mathord{}$4} SYM and
  multiparticle states of AdS(5) SUGRA''},
Nucl.~Phys.~\textbf{B643}~(2002)~49,
\hypref{http://arXiv.org/abs/hep-th/0206020}{\texttt{hep-th/0206020}}.

\end{thebibliography}
%\bibliographystyle{BMNSConf}

%%%%%%%%%%%%%%%%%%%%%%%%%%%%%%%%%%%%%%%%%%%%%%%%%%%%%%%%%%%%%%%%%

%%%%%%%%%%%%%%%%%%%%%%%%%%%%%%%%%%%%%%%%%%%%%%%%%%%%%%%%%%%%%%%%%

\end{document}